\documentclass[lettersize,journal]{IEEEtran}
\usepackage{amsmath,amsfonts}
\usepackage{algorithmic}
\usepackage{array}
\usepackage[caption=false,font=normalsize,labelfont=sf,textfont=sf]{subfig}
\usepackage{textcomp}
\usepackage{stfloats}
\usepackage{url}
\usepackage{verbatim}
\usepackage{subcaption}
\usepackage{graphicx}
\usepackage{algorithm}
\usepackage{mathtools}
\usepackage{mathrsfs}
\usepackage{xcolor}
\usepackage{mathabx}
\usepackage{pifont}% 
\usepackage{subfig}
\usepackage{amssymb}% http://ctan.org/pkg/amssymb
\usepackage{pifont}% http://ctan.org/pkg/pifont
\DeclareMathAlphabet{\mathpzc}{OT1}{pzc}{m}{it}
\usepackage[algo2e,ruled, vlined, linesnumbered]{algorithm2e}
\def\BibTeX{{\rm B\kern-.05em{\sc i\kern-.025em b}\kern-.08em
    T\kern-.1667em\lower.7ex\hbox{E}\kern-.125emX}}
\usepackage{balance}
\usepackage{hyperref}  
\usepackage{cleveref}
\newtheorem{proposition}{Proposition}

\usepackage{tikz}

\title{User-centric Flexible Resource Management Framework for LEO Satellites with Fully Regenerative Payload}
\author{ Sovit Bhandari, \IEEEmembership{Student Member, IEEE}, Thang X. Vu, \IEEEmembership{Senior Member, IEEE}, and\\ Symeon Chatzinotas, \IEEEmembership{Fellow, IEEE} 
	\thanks{The authors are with the Interdisciplinary Centre for Security, Reliability and Trust (SnT), University of Luxembourg, L-1855 Luxembourg, Luxembourg. E-mail: \{sovit.bhandari, thang.vu, symeon.chatzinotas\}@uni.lu.}
	\thanks{This work is supported by the Luxembourg National Research Fund (FNR), grant reference FNR/IPBG19/14016225/INSTRUCT and FNR/C22/IS/17220888/RUTINE. For the purpose of open access, the author has applied a Creative Commons Attribution 4.0 International (CC BY 4.0) license to any Author Accepted Manuscript version arising from this submission. Parts of this work were presented in the IEEE International Conference on Communications Workshops \cite{SovitICC23} }
}

\begin{document}

\maketitle

\begin{abstract}
The regenerative capabilities of next-generation satellite systems offer a novel approach to design low earth orbit (LEO) satellite communication systems, enabling full flexibility in bandwidth and spot beam management, power control, and onboard data processing. These advancements allow the implementation of intelligent spatial multiplexing techniques, addressing the ever-increasing demand for future broadband data traffic. Existing satellite resource management solutions, however, do not fully exploit these capabilities. To address this issue, a novel framework called flexible resource management algorithm for LEO satellites (FLARE-LEO) is proposed to jointly design bandwidth, power, and spot beam coverage optimized for the geographic distribution of users. It incorporates multi-spot beam multicasting, spatial multiplexing, caching, and handover (HO).
In particular, the spot beam coverage is optimized by using the unsupervised K-means algorithm applied to the realistic geographical user demands, followed by a proposed successive convex approximation (SCA)-based iterative algorithm for optimizing the radio resources. Furthermore, we propose two joint transmission architectures during the HO period, which jointly estimate the downlink channel state information (CSI) using deep learning and optimize the transmit power of the LEOs involved in the HO process to improve the overall system throughput. Simulations demonstrate superior performance in terms of delivery time reduction of the proposed algorithm over the existing solutions.
\end{abstract}

\section{Introduction}
In the context of satellite constellations, the LEO constellation is considered suitable for broadband services due to its small round-trip delay compared to other satellite constellations. Thanks to advanced payload technology, the LEO satellites are now seen as key enablers for the beyond 5G (B5G) and sixth-generation (6G) communications systems, as they can intelligently deliver low-cost, higher-throughput broadband services to underserved areas \cite{b1}, \cite{b2}.

The success of LEO satellites to B5G/6G relies on the operating and payload architecture. Traditionally, two main configurations have been prevalent: the wide beam and the multiple spot beam configurations. The wide-beam configuration is characterized by wide coverage and is mainly used for broadcasting applications, while the multiple-spot beams are specifically designed for broadband services \cite{b3}. These configurations excel at providing dedicated services but lack the flexibility to effectively handle dynamic and complex situations such as targeted users' mobility and time-varying demand. This lack of flexibility is favored in traditional satellite architectures due to the high cost and delays associated with the payload changes \cite{b4}. However, recent advancements in payload technologies, such as digital transparent payload (DTP) and active onboard antennas, enable efficient and reconfigurable hybrid broadcast/broadband modes \cite{b5}. 

Current DTP, however, has limited capabilities such as flexible channelization and rudimentary power control/sharing among carriers \cite{b6}. Thus, to address the shortcomings of the DTP, satellite companies are shifting their focus to incorporating advanced regenerative (fully digital) payload technology, which integrates a regenerative processor, electronically steered phased-array antennas, and optional memory units \cite{b7}. This transition allows for the optimization of various functionalities, including beamforming, spot beam coverage patterns, signal quality, bandwidth, and power as per the traffic demand \cite{b8},\cite{b9}.
To minimize overall latency and further enhance the quality of service (QoS), the regenerative payload's optional memory unit can be used for caching in LEO satellites. This approach is favored over the terrestrial networks because data cached in terrestrial networks must traverse multiple hops, which causes frequent handovers (HO) at the gateways (GWs) unless the requesting user equipment (UEs) are adjacent to edge nodes \cite{b10}. Moreover, the regenerative payload of the satellite constellation allows for the flexibility of on-demand multicasting services, potentially enabling the simultaneous delivery of cached content to different communities of users spread across different geographic areas \cite{b3}, \cite{b11}.

The successful launch of the OneWeb's LEO satellite, JoeySat, in May 2023,  funded by the European Space Agency and UK Space Agency, showcases the incorporation of the flexible software-defined regenerative processor along with multi-spot beam electronically steered phase array antennas. This implementation fulfills the demand-based beam tailoring and steering capability \cite{b12}. However, a complete package of algorithm design is needed to fully leverage the functionality of fully flexible regenerative payload-enabled satellites.

\subsection{Related Works}
Several studies have been conducted to partially exploit the flexible payload capabilities \cite{b4},\cite{b9},\cite{b13} -- \cite{b17}. 
In \cite{b9}, a bandwidth and power optimization method is proposed for non-geostationary orbit (NGSO) based on realistic demands. In \cite{b13}, a demand-driven geostationary orbit (GEO) beam steering and beam patterning method using flexible regenerative payload capabilities is proposed. The authors of \cite{b14}--\cite{b16} propose a caching policy in LEOs using flexible onboard regenerative payload capabilities to minimize content delivery delay and maximize the probability of successful delivery based on predefined beam coverage, transmit power, and operating bandwidth. In \cite{b4}, the authors aim to leverage the regenerative payload-enabled capabilities, such as digital beamforming, caching, and bandwidth optimization, considering the realistic demands. 
% This work specifically utilizes the digital beamforming technique to create pre-defined broadcast and broadband beams, enabling the transmission of cached contents from the GEO satellite to content delivery networks to minimize cache feeding time and maximize the cache hit ratio.
%
Inspired by \cite{b4}, to address the shortcomings of \cite{b14}--\cite{b16} to some extent, considering the flexible regenerative payload enabled LEOs capabilities, in \cite{b17}, an optimization problem is formulated at two different time scales to maximize the utility function in the integrated satellite-terrestrial network by considering the joint design of cache placement, multicast-beamforming, base station and satellite clustering, and transmit power. However, the optimal use of satellite operating bandwidth and spot beam coverage was not considered therein. 
% Moreover, both the terrestrial and satellite nodes transmitted the same symbol, which leads to excessive bandwidth usage. 

Since the LEO satellites can only provide uninterrupted service to the particular area in the earth fixed beam scenario for about 10 to 15 minutes during one orbital, it is crucial to consider the HO scenario via inter-satellite link (ISL) \cite{b18}. Unfortunately, the existing literature lacks proper algorithm designs for HO duration involving multiple LEOs. To fulfil this research gap and address the shortcomings of \cite{b14}--\cite{b17}, we propose a flexible resource management algorithm that fully leverages the flexible regenerative payload capabilities and efficiently utilizes the ISL during the HO periods. 
% for LEO satellite's regenerative payload exploitation (FLARE-LEO) in this paper. FLARE-LEO aims to address the existing regenerative payload capabilities and also utilizes the ISL capability of regenerative payload effectively for the HO scenarios. 

\subsection{Contributions}
In this paper, we propose FLARE-LEO, 
% (LEO satellite's regenerative payload exploitation), 
a collaborative algorithm that leverages the flexible payload and electronically steered phased array antennas embedded in LEOs. FLARE-LEO incorporates various capabilities of LEOs, including demand-based adaptive beam patterning and steering, multi-spot beam multicasting, caching, bandwidth and power optimization, as well as ISL-HO. Our contributions can be summarized as follows:

\begin{itemize}
  \item We formulate a joint design of spot beam coverage, operating bandwidth, and multi-user precoding vectors to minimize the average delivery time in LEO-assisted caching networks, including HO scenarios. Although flexible bandwidth has been considered in satellite communications, to the best of our knowledge, this is the first work exploiting spatial multiplexing \cite{b19} technique within each spot beam in LEO-enabled caching systems thanks to fully flexible regenerative payload and electronically steered phased array antennas capabilities.   
  \item We propose to solve the joint optimization problem via two sub-problems: beam coverage design and radio resource allocations. 
  % To solve the first sub-problem, we draw inspiration from benchmark clustering strategies \cite{b20}, utilizing an unsupervised K-Means++ \cite{b21} clustering technique.  
  Unlike other clustering strategies \cite{b20}, our approach guarantees non-overlapping and non-empty clusters, aligning with our goal of creating distinct spot beams and optimizing their coverage area. To tackle the non-convexity of the second sub-problem, we reformulate it using a difference-of-convex (DC) representation and propose two successive convex approximation (SCA)-based iterative algorithms for joint optimization of frequency bandwidth and multi-user precoding vectors, applied to both optimal and zero-forcing (ZF) precoding designs. It is worth noting that the solution in \cite{b21} is not applicable in our system since it does not consider the bandwidth allocation.
  \item We propose novel architectures for joint resources optimization between two LEOs during the HO period, namely \emph{centralized architecture}, in which the joint optimization is executed in the GW, and \emph{distributed architecture}, in which each LEO optimizes its own radio resources and exchanges parts of the outputs to the other via ISL. 
  These architectures differ in their computational capabilities, packet overhead, and communication needs between two LEOs. 
  In addition, a deep learning (DL)-based channel state information (CSI) prediction is proposed during the HO period to improve the effective system throughput. 
  \item Finally, the advantages of the proposed framework are demonstrated via numerical results based on the realistic Movielens dataset \cite{b22}. Simulation results indicate that the adaptive beam scenario outperforms the fixed beam scenario by at least 1.22 times in terms of effective mean data rate when the total power of the LEOs is varied between (25--35) dBW. Additionally, the effective mean data rate of the proposed design in HO periods is at least 1.5$\times$ higher the conventional method without joint transmission.
  
  % Moreover, our proposed optimal precoding approach shows a performance gain of over 41\% compared to the zero-forcing (ZF) based approach during the time without HO, when the antenna element spacing is $\lambda/2$ and the Rician factor is 10 dB. In addition, the average content delivery time is reduced by at least 74\% when the backhaul rate varies between 0.25 and 1 Gbps. These results assume that the content requested by the user device is fully stored on LEO.
  
\end{itemize}

\subsection{Organization}

The remainder of this paper is organized as follows. Section II describes the system model and parameters. Section III presents the problem formulation and proposed solution. Section IV introduces the HO scenario and DL-based CSI prediction scheme. Section V presents the different HO schemes based on computational capability and overhead. Section VI demonstrates the effectiveness of the proposed scheme using numerical results. Finally, Section VII concludes the paper.

\textit{Notations:}
The superscript $(.)^H$ stand for the
Hermitian transpose. $\lvert.\rvert$ and $\lvert\lvert.\rvert\rvert$
denote the amplitude and the $l_{2}$-norm of a set, respectively. The description of the main notations is summarized in Table I.

\begin{table}[t!]
\centering
% \color{blue}
\caption{Summary of Main Notations}
\label{table:1}
\begin{tabular}{ |p{1.5cm}||p{6.5cm}| }
 \hline
 \textbf{Symbols} & \textbf{Description} \\
 \hline
 $M$  &  Number of spot beams\\
 $N$ &   Spatial multiplexing gain in each spot beam  \\
 $G_{m}, G_u$&  Antenna gain of spot beam $m$ and UE $u$\\
 $\theta_{m},\phi_{m}$ & Elevation angle and azimuth angle\\ 
 $D_{m}, r_{m}$ &  Slant distance and radius of coverage of beam $m$ \\
$T, \tau_{slot}$, $\tau_{csi},\tau_{pro}$ &  Service duration time, time slot duration, channel estimation time, and processing time \\
$\boldsymbol{h}_{u,k,a,m}$ & Downlink channel coefficient\\
$\boldsymbol{w}_{k,a,m}$ & Precoding vector for the multicast group $k$ of AUG $a$\\
$\check{I}_{m}, \hat{I}_{m}$ & Intra-beam and inter-beam interferences\\
$b_{a,m}$ & Bandwidth of associated group $a$ within spot beam $m$    \\
$\mathcal{L}_\mathpzc{d}(\Theta)$& CNN-CSI loss function \\
\hline
\end{tabular}
\vspace{-0.3cm}
\end{table}

\section{System Model}

\begin{figure}[b!]
\vspace{-0.5cm}
\centerline{\includegraphics[width=0.49\textwidth]{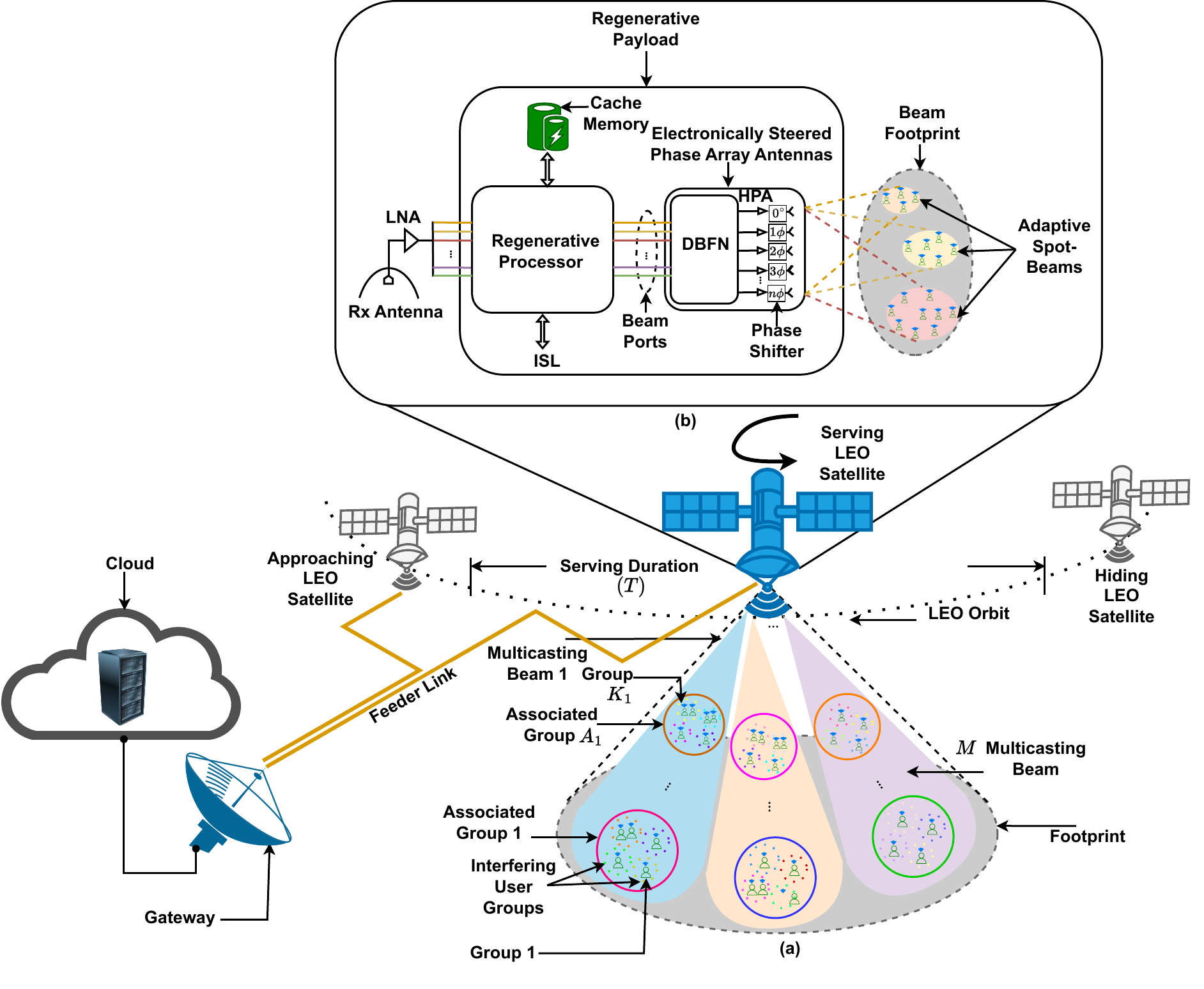}}
\caption{A downlink multicasting LEO satellite system with a regenerative payload architecture. (a) Multicasting LEO satellite system. (b). Regenerative payload architecture.}
\label{fig1}
\end{figure}

We consider a LEO constellation providing services in a given area, in which a LEO satellite is serving the users at a given time. HO occurs when the serving LEO satellite is departing and a new LEO is joining the area. Without loss of generality, the considered system not in the HO period comprises a LEO satellite serving a set $\mathcal{U} = \{1,2,...,u,...,U\}$ of $U$ single-antenna UEs within its coverage, a GW, and a centralized cloud, as shown in Fig. 1.  The operation during the HO period will be presented in Section IV. 
% 
% \textcolor{blue}{The impact of imperfect CSI is beyond the scope of this work and the readers are referred to \cite{b23} for information on the impact of channel estimation errors and outdated CSI for the satellite standard DVB-S2(X).}
%
The LEO satellite is equipped with a cache-enabled flexible regenerative payload and electronically steered phased array antennas that can generate $M$ spot beams of arbitrary shapes \cite{b23}
% , thanks to the digital beamforming (BF) in phased array antennas, 
to adaptively serve UEs within its footprint. 
% This digital beamforming technique effectively optimizes the energy consumption of high-power amplifiers (HPAs).
For ease of analysis, we assume the shape of spot beam $m$ to be circular and 
to leverage the users clustering and spot beam optimization, we assume a uniform antenna radiation pattern within a spot beam. It is worth noting that once the spot beams are determined, actual location-dependent path losses are employed to compute the received signal power. The spot beam gain $G_m (\theta_m, \phi_m)$ can be computed as \cite{b24}:
\begin{equation} \label{eq1}
\small
\begin{split}
\small
G_m (\theta_m, \phi_m) & = \frac{\text{Area of isotropic sphere}}{\text{Area of spot beam rad. pattn.}} = \frac{4 D_m^2}{ r_m^2} , \, \forall m, \\
\end{split}
\end{equation}
where $\theta_m$ and $\phi_m$ are the elevation and azimuth angles in radians relative to the boresight of the spot beam $m$, $r_m$ is the spot beam's radius, and $D_{m}$ is the slant distance between the LEO satellite and the $m$-th spot beam. 
To efficiently serve the UEs over a geographical area, the spot beams are designed to be non-overlapping and adequately spaced, which allows full-frequency reuse. Thanks to the advanced payload technology, each LEO satellite can deliver up to $N$ spatial multiplexing data streams in each spot beam \cite{b25}.

%%%%%%%%%%%%%%%%%%%%%%%
\subsection{Caching Model}
 By equipping with the advanced flexible regenerative payload, the LEO satellite is able to process data and has a limited cache memory of $C$ bits. The $U$ UEs are interested in the content library of $\mathcal{F} = \{1,2,...,f,..., F\}$ at the centralized cloud consisting of $F$ files. Due to the non-geostationary nature, the LEO satellite has a limited service duration of $T$ for each satellite pass of a considered area.
 %
 % , which is divided into two phases: cache placement phase of $\tau$ c.u. and the delivery phase of $T - \tau$ c.u. Due to the fast movement of the LEO satellite, elevation and azimuth angles, hence the channel gains, largely vary during each service duration. \textcolor{blue}{Therefore, the delivery phase is considered an aggregation of multiple time slots, each of duration $t$ c.u., during which the channels are assumed to be quasi-static. To minimize the operating cost, the spot beams' coverages are optimized for the whole service duration of the LEO satellite.}
 %
 We consider offline caching policy \cite{b26} in which the demand vector $\boldsymbol{L}$ is obtained in advance, e.g., via historical average or prediction model \cite{b27}. %$\footnote[1] {The joint design for on-line caching setting is left for future work.}$.
 Based on $\boldsymbol{L}$, the cache placement is executed at the beginning of each service duration based on generic caching models, such as most popular caching (MPC), uniform caching (UC), and random caching (RC). We focus on the transmission design in the delivery phase. 

\subsection{User Grouping}
\label{sec:user-grouping}
 To exploit the flexible multi-beam capability, the users are served in groups depending on their geographical locations and requested contents. Denote the set of UEs in each spot beam $m$ as $\mathcal{U}_m \subseteq \mathcal{U}$, which is further divided into $K_m$ groups. The users within the same group request the same content file. 
 % groups based on their set of requested contents $\mathcal{F}_{m} = \{1,2,...,f,...,F_{m}\}$. 
 % UEs within the same spot beam requesting the same file are assigned to the same group. 
 If $K_m \leq N$, all user groups can be served simultaneously using the whole bandwidth $B$ via spatial multiplexing techniques, i.e., multi-user precoding. Otherwise, $K_m$ groups are divided into $A_m = \lceil\frac{K_m}{N}\rceil$ associate user groups (AUGs). Different AUGs are served via orthogonal frequency bandwidths, while the users within one AUG are served simultaneously via multi-user precoding technique. Let $\mathcal{A}_{m}=\{a_{1}^{m},a_{2}^{m},...,a_{A_m}^{m}\}$ denote the set of $A_m$ AUGs in spot beam $m$, and $\mathcal{K}_{a,m}$ denote the set of users belong to the AUG $a$ of spot beam $m$.

To illustrate, suppose 8 users $\{u_{1}, u_{2},..., u_{8}\}$ within spot beam $m$ requests the corresponding files $\{f_{1}, f_{2}, f_{3}, f_{3}, f_{4}, f_{5}, f_{5}, f_{6}\}$, which includes $K_m=6$ distinct files. If $N = 4$, then there are  \emph{two AUGs, which are: 
 $\mathcal{K}_{1,m}=\{u_1, u_2, u_3, u_4, u_5\}$ and $\mathcal{K}_{2,m}=\{u_6, u_7, u_8\}$}\footnote[1]{
This is one of possible AUG partitions. Optimal user grouping is not considered in this work.}.

\subsection{Transmission Model }
We focus on the signal transmission during the delivery phase in which the LEO satellite serves the users' requested contents. The service duration is divided into multiple time slots, whose duration is determined by the channel coherence time. The satellite-user channels are assumed quasi-static within one time slot and vary from one time slot to another. For a particular time slot, the signal received by the UE $u$ in group $k$ of AUG $a$ ($k \in \mathcal{K}_{a,m}$) in spot beam $m$ not in the HO time (see Section~\ref{sec:HO} for HO transmission) can be written as: 
\begin{align}
\small
y_{u,k,a,m}=&\boldsymbol{h}_{u,k,a,m}^{ H}\boldsymbol{w}_{k,a,m}{s}_{k,a,m}+ \check{I}_{m} + \hat{I}_{m}  + n_{u}, 
\end{align}
where $\boldsymbol{h}_{u,k,a,m} \in \mathbb{C}^{N\times1}$ is the downlink channel coefficient to UE $u$ of multicast group $k$ of AUG $a$; $\boldsymbol{w}_{k,a,m}\in\mathbb{C}^{N\times1}$ is the precoding vector designed for the multicast group $k$ of AUG $a$; ${s}_{k,a,m} \in \mathbb{C}$ is the data symbol requested by UEs of group $k$ of AUG $a$ via multicast spot beam $m$ with $\mathop{\mathbb{E}}[|{s}_{k,a,m}|^2]=1$; 
% $\check{I}_{m}$ and $\hat{I}_{m}$ are the intra-spot beam interference and inter-spot beam interference, respectively;  
and $n_{u} \sim$ $\mathcal{CN}(0,\sigma_u^2)$ is the additive white gaussian noise (AWGN). $\check{I}_{m}$ is intra-spot beam interference that  is caused by the concurrent transmission to different user groups within the same AUG $a$ of spot beam $m$, and $\hat{I}_{m}$ represents inter-spot beam interference caused by power leakage from the adjacent beams. They are computed as follows:
\begin{align}
\small
\check{I}_{m}\triangleq \sum_{k'\in\mathcal{K}_{a,m} \symbol{92}\{k\}}{\boldsymbol{h}_{u,k,a,m}^{H}\boldsymbol{w}_{ k',a,m}s_{k',a,m}}; \ 
{}\hat{I}_{m} \triangleq \eta_{m} \frac{P_{{\sum}_{m'}}}{B}b_{a,m} , \notag
\end{align}
where $\eta_m$ is the aggregate of the $m$-{th} spot beam inter-spot beam attenuation factor and the free-space path loss, $P_{{\sum}_{m'}}/B$ represents the accumulated interference density caused by the adjacent spot beams, and $b_{a,m}$ is the frequency bandwidth allocated to AUG $a$ within spot beam $m$. The typical value for the inter-spot beam attenuation factor is around -30 dB. After the Doppler compensation, we have 
 ${\boldsymbol{h}}_{u,k,a,m} = {g}_{u} \boldsymbol{v}_{u}(\varphi_u)$, where ${g}_{u}$ is the channel gain and $\boldsymbol{v}_{u}(\varphi_u) \in \mathbb{C}^{N\times1}$ is the downlink array response vector for UE $u$, wherein $\varphi_u$ is the angle of departure (AoD) \cite{b28}. 
 
 We assume a Rician fading channel $g_u = \sqrt{\beta_u}\hat{g}_u$, where $\beta_u = G_m G_u M \lambda^2 /(4\pi D_u)^2$ is the large-scale fading and $\hat{g}_u = \alpha_u({h}_{LoS}\sqrt[]{{\kappa_u}/{(\kappa_u+1)}}+{h}_{NLoS}\sqrt[]{{1}/{(\kappa_u+1)}})$ denotes the small-scale fading channel model, with $\kappa_u$ represents the Rician factor, $\alpha_u=\mathop{\mathbb{E}}\{ |\hat{g}_u|^2\}$,
  % wherein $G_m, G_u$ are transmit and receive antenna gains, respectively, and $D_{u}$ is the slant distance between LEO satellite and UE.  $\hat{g}_u = \alpha_u({h}_{LoS}\sqrt[]{{\kappa_u}/{(\kappa_u+1)}}+{h}_{NLoS}\sqrt[]{{1}/{(\kappa_u+1)}})$ denotes the small-scale fading channel model, wherein $\kappa_u$ represents the Rician factor, $\alpha_u=\mathop{\mathbb{E}}\{ |\hat{g}_u|^2\}$, 
 ${h}_{LoS}$ is the deterministic line-of-sight (LoS) part, and ${h}_{NLoS}$ represents the non-LoS (NLoS) component. Other parameters are given in Table~\ref{table:1}.  The real and imaginary parts of $\hat{g}_{u}$ are independently and identically distributed as $\mathcal{N}$$\mathbb ( \sqrt[]{\kappa_u \alpha_u/2 (\kappa_u+1)}$, $\alpha_u/2(\kappa_u+1))$. Assuming perfect CSI at the satellite, the signal-to-interference-plus-noise ratio (SINR) of UE $u$ of AUG $a$ at spot beam $m$ is calculated as follows:
 \begin{equation}\label{eq5}
 \small
{\gamma_{u,k,a,m}=\frac{|\boldsymbol{h}_{u,k,a,m}^{H}\boldsymbol{w}_{k,a,m}|^2}{\sum_{k'\in\mathcal{K}_{a,m} \symbol{92}\{k\}}|{\boldsymbol{h}_{u,k,a,m}^{H}\boldsymbol{w}_{k',a,m}}|^2+ \hat{I}_{agg}b_{a,m}}},  
\end{equation}
where $\hat{I}_{agg} \triangleq (\eta_m (P_{\sum_{m'}}/{B}) + N_{0})$ and $N_{0}$ is the noise spectral density. The impact of imperfect CSI is studied in Section VI-F.

The effective transmission rate of a group $k$ within AUG $a$, determined by the weakest users in the group, in spot beam $m$, is calculated as follows:
\begin{equation}\label{eq6}
R_{k,a,m} = \Phi b_{a,m} \log_2(1+{\text{min}_{u}\{\gamma_{u,k,a,m}\}}). 
\end{equation}
where $\Phi \triangleq 1 - \frac{\tau_{csi}+\tau_{pro}}{\tau_{slot}}$ accounts for the effective transmission time, $\tau_{slot}$ is the time slot duration, $\tau_{csi}$ is the max channel estimation time, and $\tau_{pro}$ is the processing time whose value largely depends upon the beamforming techniques and the hardware capability of the regenerative payload.

\section{Problem Formulation and Proposed Solution}

\subsection {Problem Formulation}
In the pursuit of full exploitation of the flexible payload, we aim to jointly design the spot beam coverage $r_m$, frequency bandwidth allocation, and precoding vectors towards minimizing the worst-case average delivery latency. The joint optimization problem is formulated as follows:
\begin{subequations}\label{eq7}
\small
\begin{align}
\begin{split}
\mathcal{P}: {}& \underset{\{ \boldsymbol{w}, \boldsymbol{b},\boldsymbol{r}, \boldsymbol{K}\}}{\min}  \quad t(\boldsymbol{w}, \boldsymbol{b}, \boldsymbol{r}, \boldsymbol{K}) \label{eq7a}
\end{split} \\
\text{s.t.} {}& \quad R_{k,a,m} \ge R_{req}, \quad  \forall k,a,m, \label{eq7b} \\
 {}& \quad \sideset{}{_{k \in \mathcal{K}_{a,m}}}\sum{ \sideset{}{_{a \in \mathcal{A}_{m}}}\sum \lVert\boldsymbol{w}_{k,a,m}  \rVert^{2} \le {P_{\sum}(K_{m}}/{K})},  \forall m, \label{eq7c}\\
{}& \quad {\sideset{}{_{a\in \mathcal{A}_{m}}}\sum b_{a,m}} \le B, \quad \forall m, \label{eq7d} \\
{}& \quad {\sideset{}{_{m \in \mathcal{M}}} \bigsqcup\limits \pi r_m^2} \ge A_{\sum} \label{eq7e},
\end{align}
\end{subequations}
where $\boldsymbol{w} \triangleq \{\boldsymbol{w}_{k,a,m}\}_{\forall k,a,m}$, $\boldsymbol{b} \triangleq \{b_{a,m}\}_{\forall a,m}$, $\boldsymbol{r} \triangleq \{r_{m}\}_{m=1}^{M}$, and $\boldsymbol{K} \triangleq \{K_{m}\}_{ m=1}^{M}$ are the short-hand notations;
$R_{req}$ is the minimum QoS requirement; $\mathcal{A}_{m}$ is the set of AUGs in spot beam $m$; $P_{\sum}$ is the total transmit power of LEO satellite; $K=\sum_{m=1}^{M}K_m$; and $A_{\sum}$ is the total service area of the LEO satellite.

 The objective function $t(\boldsymbol{w}, \boldsymbol{b}, \boldsymbol{r}, \boldsymbol{K})$ of problem \eqref{eq7} is the end-to-end transmission latency, assuming the FastForward capability \cite{b29}, is computed as follows:
\begin{equation}
\small
{t(\boldsymbol{w}, \boldsymbol{b}, \boldsymbol{r}, \boldsymbol{K}) = \sideset{}{_{\{k,a,m\}}}\max \left( \max \left(\left( {q_k}/{R_{\mathtt{X}}}+{D_k}/{c} \right),\Pi_{k}\right)\right) \label{eq8}}
\end{equation}
where $\mathtt{X} \triangleq \{k,a,m\}$ is the short-hand indexes, $q_{k}$ is the file size, $D_k = \max_{u}(D_u)$ is the slant distance between LEO satellite and $k$-th group, $c$ is the speed of light. In \eqref{eq8}, $\frac{q_{k}}{R_{\mathtt{X}}}$ and $\frac{D_{k}}{c}$ are the transmission and the propagation delays, respectively, incurred while sending files from the LEO satellite to UEs of group $k$; and $\Pi_{k}$ $\triangleq$ $\frac{(1-\mu_k)q_k}{R_{BH}}+\frac{D_0}{c}$ is the transmission and propagation delay accured in the backhaul link when sending the uncached file parts from the centralized cloud to the LEO satellite, where $\mu_{k} \in [0,1]$ denotes the fraction of the $k$-th file on LEO satellite, $R_{BH}$ is the backhaul transmission rate, and $D_0$ is the slant distance between the GW and the LEO satellite.

In problem $\mathcal{P}$, constraint \eqref{eq7b} guarantees the minimum users' QoS requirement; constraint \eqref{eq7c} limits the power allocated to each LEO satellite spot beam; constraint \eqref{eq7d} sets the total bandwidth at spot beam $m$ not exceeding $B$. Finally,  constraint \eqref{eq7e} ensures that the non-overlapping union of the coverage areas of a total number of spot beams covers at least the total service area of LEO satellite.

\textit{Difficulty to solve problem $\mathcal{P}$:} The challenge in solving problem $\mathcal{P}$ lies in both the non-convexity of the objective function and constraints \eqref{eq7b} and \eqref{eq7e}, which result in a non-deterministic polynomial time hard problem. In particular, the spot beam coverage partition does not only affect the user grouping but also the antenna radiation patterns and hence the effective channel gains. 

\subsection {Proposed Solution}
One might optimize the spot beam coverages jointly with the bandwidth and precoding vectors for every time slot. This method, however, imposes significant computation and operating costs. Instead, we design the spot beam coverages for the whole service duration $T$ and \emph{decouple the original problem $\mathcal{P}$}\footnote[2]{
It is only efficient when the geographical distribution of users and requested contents changes at a much slower rate than the time slot duration.} into two sub-problems: one optimizes the spot beam coverage area for the long-time scale $T$ and the other optimizes bandwidth and precoding vectors for the short-time scale, e.g., on a time slot basis.

% To transform the problem $\mathcal{P}$ into a more tractable form, the constraint \eqref{eq7e} is separated from the rest of the problem as the spot beam coverage area is fixed prior to the initiation of the data transmission process. Therefore, the problem $\mathcal{P}$ is decomposed into two sub-problems as:
% \emph{decouple}
%
%
%
\subsubsection{Minimization of the Spot Beam Coverage Area}
% In the conventional payload technology, e.g., SES-17, the beam coverage are predefined and fixed. The advantage of such configuration is the minimum system configuration management, however, it is unable to adapt with dynamic user demands. 
Unlike conventional payload, the full-digital payload offers full flexibility to design spot beam shapes optimized to the geographical users distribution. Since the effective channel gain is inversely proportional to the spot beam coverage, we aim to minimize the total multi-spot beams coverage while guaranteeing all the users are within the LEO satellite's coverage. The multi-spot beam coverage design is formulated as follows:
\begin{subequations}\label{eq9}
\begin{align}
\mathcal{P}_1:  {} {\min}_{\{r_m\}} &~~{\sum}_{m \in \mathcal{M}} \pi r_m^2 \label{eq9a} \\ 
\text{s.t.} {}& ~~ \sideset{}{_{m \in \mathcal{M}}} {\bigsqcup}{\pi r_m^2} \ge A_{\sum}, \quad  \forall m,  \label{eq9b}\\
&~~ 0 < r_m \leq r_{\text{Max}}, \forall m, \label{eq9c}
\end{align}
\end{subequations}
where $r_{\text{Max}}$ is the maximum spot beam radius. 

Intuitively, problem $\mathcal{P}_1$ aims at finding the optimal radius of $M$ non-overlapping spot beams, while ensuring that all the users are within the coverage of the designed spot beams, as stated in constraint \eqref{eq9b}.
To solve the problem $\mathcal{P}_1$, we employ the K-Means++ \cite{b30} clustering technique. The clustering is done based on the position of $U$ UEs that demand the service, so the spot beam center is likely to point in the direction where the number of UEs is dominant. Since there are $M$ spot beams, the $U$ UEs are categorized into $M$ clusters such that $m$-th spot beam serves $m$-th cluster. 
The problem $\mathcal{P}_1$ can be reformulated in terms of clustering as follows:
\begin{subequations} \label{eq10}
\begin{align}
\mathcal{P}_{1}^{'} : {} &\sideset{}{_{\{ \mathcal{U}_m,\boldsymbol{c}_{\mathcal{U}_m},r_{m}\}}} \min ~~ \sideset{}{_{m \in \mathcal{M}}}\sum\pi r_m^2 \label{eq10a} \\ 
\text{s.t.} {}& ~~ \eqref{eq9c};~ \sideset{}{_{m \in \mathcal{M}}}\bigsqcup\mathcal{U}_m ==  \mathcal{U}, \label{eq10b} \\ 
{}& ~~ (\boldsymbol{y}_u - \boldsymbol{c}_{\mathcal{U}_m}) (\boldsymbol{y}_u - \boldsymbol{c}_{\mathcal{U}_m})'\le r_m^2, \quad  \forall u,m, \label{eq10c}
\end{align}
\end{subequations}
where $\boldsymbol{y}_u$ is the 2-D coordinate of user $u$, $\mathcal{U}_m$ is the set of UEs in the $m$-th cluster and $\boldsymbol{c}_{\mathcal{U}_m}$ is the 2D centroid of $m$-th cluster. Constraint \eqref{eq10b} ensures that all unique UEs lie within the total service area of LEO satellite; constraint \eqref{eq10c} guarantees that UEs are clustered based on the Euclidean distance between $\boldsymbol{y}_{u}$ and $\boldsymbol{c}_{\mathcal{U}_m}$, which is bounded by the radius of coverage of the cluster, i.e., $r_m$.

The procedure to obtain $\mathcal{U}_m$, $\boldsymbol{c}_{\mathcal{U}_m}$, and $r_m$ is shown in Algorithm~\ref{algorithm:1}. To find the boundaries of the clusters, Voronoi tessellation technique \cite{b31} is used, where the boundaries of the Voronoi polygons are computed using $\boldsymbol{c}_{\mathcal{U}_m}$. However, for mathematical tractability, the coverage area of the spot beam is considered circular.
Using the outputs of Algorithm~\ref{algorithm:1}, user grouping is done as shown in Section~\ref{sec:user-grouping} to get $\mathcal{A}_{m}$, $\mathcal{K}_{a,m}$, and $K_m$, which are used in solving the second sub-problem.

 %%%%%%%%%%%%%%%%%%%%%%%%%%%%%%%%%%%%%%%%%
\subsubsection{Minimization of Content Delivery}
\label{sec:jointframework}
Once the spot beams are determined, we are ready to optimize the bandwidth allocation and precoding vectors to minimize content delivery latency. We assume that the time slot duration is sufficient for the satellite to serve the current users' requests, and the joint bandwidth and precoding vectors design is formulated as follows:
\begin{subequations}\label{eq11}
\begin{align}
% \mathcal{P}_2:& 
% \min_{\substack{\{\boldsymbol{w}, \boldsymbol{b}\}}}\! \max_{\{\mathtt{k,a,m}\}} \resizebox{.73\hsize}{!}{\left(\! \max \left(\!\frac{q_k}{{b}_{a,m} \Phi\log_2(1+ \text{min}_{u}\{\gamma_{u,\mathtt{X}}\})} \!+\! \frac{D_k}{c}, \Pi_k\! \right)\! \right) 
\mathcal{P}_2:&~ {\min}_{\boldsymbol{b}, \boldsymbol{w}} ~~t(\boldsymbol{w}, \boldsymbol{b}, \boldsymbol{r}, \boldsymbol{K}) \label{eq11a} \\
\text{s.t.}{}& ~~ \eqref{eq7c};\ \eqref{eq7d};\ b_{a,m}\Phi\log_2(1 + \min_u\{\gamma_{u,k,a,m}\}) \ge \notag \\
& \qquad\qquad\qquad  R_{req},  \forall \, k,a,m, \label{eq11b} 
\end{align}
\end{subequations}
where $t(\boldsymbol{w}, \boldsymbol{b}, \boldsymbol{r}, \boldsymbol{K})$ is given in \eqref{eq8}. 
% The objective function \eqref{eq11a} is obtained by replacing the values of \eqref{eq8} and \eqref{eq6} in \eqref{eq7a}, and the \eqref{eq11b} is obtained by replacing the value of \eqref{eq6} in \eqref{eq7b}.

The problem $\mathcal{P}_2$ is non-convex due to the objective function and the constraint \eqref{eq11b}. To tackle this difficulty, we introduce slack variables $z_{k,a,m}$, $\gamma_{k,a,m}$ and reformulate $\mathcal{P}_2$ into a more tractable form as follows:
% Thus, to solve problem $\mathcal{P}_2$, following the relaxation method, the minimal operators $\min_u\{\gamma_{u,k,a,m}\}$ in the objective function is replaced by positive auxiliary variable $\gamma_{k,a,m}$. Also, $b_{a,m}\log_2(1+{\text{min}_u\{\gamma_{u,k,a,m}\}})$ in the objective function of problem $\mathcal{P}_2$ is replaced by the positive auxiliary variable $z_{k,a,m}$ to transform the problem $\mathcal{P}_2$ into a more tractable form. As a result, the problem $\mathcal{P}_2$ can be equivalently reformulated as:
%
\begin{subequations}\label{eq12}
\small
\begin{align}
% \begin{split}
\small
\mathcal{P}_{2}^{'}:   &   \min_{\substack{\{\boldsymbol{w}, \boldsymbol{b}, \boldsymbol{\gamma}, \boldsymbol{z}\}}} \max_{k,a,m} \big( \max \big(\frac{q_k}{z_{k,a,m}}+\frac{D_k}{c}, \Pi_k \big)\big)\label{eq12a}\\
% \end{split}\\
\text{s.t.} & ~ b_{a,m} \Phi \log_2(1+\gamma_{k,a,m}) \ge z_{k,a,m},  \forall k,a,m, \label{eq12b} \\
\begin{split}
%& \, \resizebox{.80\hsize}{!}{( \lvert \boldsymbol{h}_{u,k,a,m}^{H}\boldsymbol{w}_{k,a,m}\rvert^2)/(\sideset{}{_{k'\in\mathcal{K}_{a,m} \symbol{92}k}}\sum \lvert{\boldsymbol{h}_{u,k,a,m}^{H}\boldsymbol{w}_{k',a,m}}\rvert^2} \\ 
& \, (\lvert \boldsymbol{h}_{u,k,a,m}^{H}\boldsymbol{w}_{k,a,m}\rvert^2)/({\sum}_{k'\in\mathcal{K}_{a,m} \symbol{92}k} \lvert{\boldsymbol{h}_{u,k,a,m}^{H}\boldsymbol{w}_{k',a,m}}\rvert^2 \\ 
& \quad +\hat{I}_{agg}b_{a,m})  \ge \gamma_{k,a,m} ,   \quad \forall u,k,a,m,
\end{split} \label{eq12c} \\
& ~ b_{a,m}\Phi \log_2(1+\gamma_{k,a,m}) \ge R_{req}, \quad  \forall k,a,m,  \label{eq12d}\\
& ~ \eqref{eq7c},\eqref{eq7d}, \nonumber
\end{align}
\end{subequations}
where $\boldsymbol{\gamma} \triangleq \{\gamma_{k,a,m}\}_{\forall k,a,m}$ and $\boldsymbol{z} \triangleq \{z_{k,a,m}\}_{\forall k,a,m}$.

The main challenge in solving problem $\mathcal{P}_2^{'}$ lies in the first three constraints, i.e., \eqref{eq12b}, \eqref{eq12c}, and \eqref{eq12d}. We can handle constraint \eqref{eq12b} by considering the slack variable $x_{k,a,m}$, which can be reformulated as:
\begin{gather}
\small
\Phi\log_2(1+\gamma_{k,a,m}) \ge x_{k,a,m}, \label{eq13} \\
b_{a,m}x_{k,a,m} \ge z_{k,a,m}. \label{eq14} 
\end{gather}
Constraint \eqref{eq13} is convex, and to deal with constraint \eqref{eq14}, we use an equivalent representation as:
\begin{equation}\label{eq15} 
\small
\eqref{eq14} \Leftrightarrow
(b_{a,m}+x_{k,a,m})^2 \ge 2z_{k,a,m} + b_{a,m}^2 + x_{k,a,m}^2,  
\end{equation}
 which has a difference-of-convex (DC) form as both sides are convex functions. The DC programming in constraint \eqref{eq15} can be easily tackled using the iterative-based SCA method by taking the first-order approximation of the left-hand-side (LHS) of the constraint \eqref{eq15}. Let ${\bar{b}}_{a,m}$ and ${\bar{x}}_{k,a,m}$ be the feasible values of the constraint \eqref{eq15} in the current iteration. In the next iteration, the constraint \eqref{eq15} can be approximated as a convex constraint as:
\begin{equation}\label{eq16}
\small
\begin{split}
{}&2(b_{a,m}+x_{k,a,m})(\bar{b}_{a,m}+\bar{x}_{k,a,m})-(\bar{b}_{a,m}+\bar{x}_{k,a,m})^2 \ge  \\ {}& \qquad \qquad \qquad \qquad \qquad 2z_{k,a,m} + b_{a,m}^2 + x_{k,a,m}^2.
\end{split}
\end{equation}
To tackle the non-convexity of constraint \eqref{eq12c}, we represent it in an equivalent form as:
\begin{align}\label{eq17}
\small
{}&
(\lvert\boldsymbol{h}_{u,k,a,m}^{H}\boldsymbol{w}_{k,a,m}\rvert^2)/\gamma_{k,a,m} \ge \nonumber \\ {}& {\sideset{}{_{k'\in\mathcal{K}_{a,m} \symbol{92}\{k\}}}\sum\lvert{\boldsymbol{h}_{u,k,a,m}^{H}\boldsymbol{w}_{k',a,m}}\rvert^2 +  \hat{I}_{agg}b_{a,m}}. 
\end{align}
 Since the constraint \eqref{eq17} is also in DC form, we use the SCA method to solve it iteratively. Taking ${\bar{w}}_{k,a,m}$ and ${\bar{\gamma}}_{k,a,m}$ as the feasible value, \eqref{eq17} can be approximated as:
\begin{align}\label{eq18}
\small
&\dfrac{2\boldsymbol{w}_{k,a,m}^{H}\boldsymbol{H}_{u,k,a,m}\bar{\boldsymbol{w}}_{k,a,m}}{\bar{\gamma}_{k,a,m}}-{{\gamma}_{k,a,m}}\dfrac{\bar{\boldsymbol{w}}_{k,a,m}\boldsymbol{H}_{u,k,a,m}\bar{\boldsymbol{w}}_{k,a,m}}{\bar{\gamma}_{k,a,m}^2}\nonumber\ge \\
&~ {\sum}_{k'\in\mathcal{K}_{a,m} \symbol{92}k}{\boldsymbol{w}_{k',a,m}^H\boldsymbol{H}_{u,k,a,m}\boldsymbol{w}_{k',a,m}}+\hat{I}_{agg}b_{a,m}, 
\end{align}
where $\boldsymbol{H}_{u,k,a,m} \triangleq \boldsymbol{h}_{u,k,a,m} \boldsymbol{h}_{u,k,a,m}^{H}$.
\begin{figure}[t!]
  \begin{minipage}[t]{0.48\textwidth}
    \begin{algorithm}[H]
      \caption{Iterative Alg. to Solve  \eqref{eq10a}}
      \begin{algorithmic}[1]
      \label{algorithm:1}
        \renewcommand{\algorithmicrequire}{\textbf{Input:}}
        \renewcommand{\algorithmicensure}{\textbf{Output:}}
        \REQUIRE $U$, $A_{\sum}$, $\boldsymbol{y}_{u}$, $M$
        \ENSURE  $\mathcal{U}_m$, $\boldsymbol{c}_{\mathcal{U}_m}$, $r_{m}$ 
        \\ \textit{Init} : \{$\boldsymbol{c}_{\mathcal{U}_m}\}_{m=1}^{M}$, $i = 1$, $I_{max}$, $dis = []$, $\epsilon$, $err = 1$
        \STATE Based on $\boldsymbol{y}_{u}$, apply the K-Means++ cluster Alg.
        \WHILE {$err > \epsilon$ and $i<I_{max}$}
        \STATE  Calculate $\mathcal{U}_m$ using K-Means++ Alg.
        \STATE Compute $\boldsymbol{c}_{\mathcal{U}_m}^{(i)}$ = \text{mean}\{${\boldsymbol{y}_u}$ $\mid$ $u$ $\in$ $\mathcal{U}_m$\} 
        \STATE Compute $err = |{\boldsymbol{c}_{\mathcal{U}_m}^{(i)}-\boldsymbol{c}_{\mathcal{U}_m}^{(i-1)}}|$
        \STATE Update $\boldsymbol{c}_{\mathcal{U}_m}^{(i-1)} \leftarrow {c}_{\mathcal{U}_m}^{(i)}$; $i\leftarrow i+1$
        \ENDWHILE
        \STATE Calculate radius $r_m$ of the cluster $\mathcal{U}_m$
        \FOR {$m = 1$ to $M$}
        \FOR {$j = 1$ to length\{$\mathcal{U}_m$\}} 
          \STATE Compute \\
          $dis^{(j)} = $ $ \sqrt{(\boldsymbol{y}_u  - \boldsymbol{c}_{\mathcal{U}_m}) (\boldsymbol{y}_u  - \boldsymbol{c}_{\mathcal{U}_m})'}$
        \ENDFOR
        \STATE Compute $r_m$ = $\lceil{(\text{max}\{dis\})}\rceil$
        \ENDFOR
      \end{algorithmic}
    \end{algorithm}
  \end{minipage}%
  \hfill%
  \end{figure}
Using $\eqref{eq16}$ and $\eqref{eq18}$, the problem $\mathcal{P}_2^{'}$ can be approximated by a convex optimization problem $\mathcal{P}_2^{''}$ as:
\begin{subequations}\label{eq19}
\begin{align}
\small
\begin{split}
\mathcal{P}_{2}^{''}\substack{({\boldsymbol{\bar{w}}},{\boldsymbol{\bar{b}}},{\boldsymbol{\bar{x}}}, {\boldsymbol{\bar{\gamma}}})}: {}&  \min_{\substack{\{\boldsymbol{w}, \boldsymbol{b}, \boldsymbol{\gamma},\boldsymbol{x},\boldsymbol{z}\}}}  \max_{k,a,m} \left( \max \left(\frac{q_k}{z_{k,a,m}}+\frac{D_k}{c}, \Pi_{k} \right)\right)\label{eq19a}
\end{split}\\ 
\text{s.t.} {}& \,  \eqref{eq7c},\eqref{eq7d},\eqref{eq13},\eqref{eq16},\eqref{eq18},\nonumber\\
 {}& \, \Phi\log_2(1+\gamma_{k,a,m}) \ge \frac{R_{req}}{b_{a,m}} \,  \forall k,a,m, \label{eq19b}
\end{align}
\end{subequations}
where $\boldsymbol{x} \triangleq \{x_{k,a,m}\}_{\forall k,a,m}$ and  \eqref{eq19b} is directly obtained from \eqref{eq12d}.

The problem $\mathcal{P}_2^{''}$ is a convex problem and it can be solved directly using the interior point method \cite{b32}. Since the solutions of problem $\mathcal{P}_2^{''}$ should satisfy all the constraints of problem $\mathcal{P}_2$, the solution provided by problem $\mathcal{P}_2^{''}$ is sub-optimal for problem $\mathcal{P}_2$ and also depends largely on the initialization of the parameters ${\boldsymbol{\bar{w}}},{\boldsymbol{\bar{b}}},{\boldsymbol{\bar{x}}},$ and  ${\boldsymbol{\bar{\gamma}}}$. Therefore, we propose Algorithm~\ref{algorithm:2} to solve \eqref{eq11}.

\subsection{Complexity of the Proposed Algorithm}
The computational complexity of Algorithm~\ref{algorithm:1} is $\mathcal{O}(2MUI_{max} + MK_{m})$ \cite{b33}. Assuming that the interior point method is used
to solve the convex problem $\eqref{eq8}$, in the worst case the
complexity is equal to the cube of the number of real variables \cite{b32}. Since there are $\left\lceil{ {K_m}/{N}}\right\rceil N^2 + \left\lceil{ {K_m}/{N}}\right\rceil N$ real variables in
the problem $\eqref{eq8}$, the complexity for solving $\eqref{eq8}$ is $\mathcal{O}\big(MI_{max}\left(\left\lceil{{K_m}/{N}}\right\rceil N^2 + \left\lceil{{K_m}/{N}}\right\rceil N\right)^{3}\big)$.
  \begin{figure}
  \begin{minipage}[h!]{0.48\textwidth}
  \begin{algorithm}[H]
  \caption{Iterative Alg. to Solve  \eqref{eq11a}}
  \label{algorithm:2}
  \begin{algorithmic}[1]
  \renewcommand{\algorithmicrequire}{\textbf{Input:}}
  \renewcommand{\algorithmicensure}{\textbf{Output:}}
  \REQUIRE $\mathcal{A}_{m}$, $\mathcal{K}_{a,m}$, $K_m$, $\boldsymbol{h}_{u,k,a,m}$, $\mu_k$, $D_k$, $c$, $R_{BH}$, $\eta_{m}$, $P_{{\sum}_{m'}}/{B}$
  \ENSURE $\boldsymbol{w}_{\scriptscriptstyle k,a,m}^{*}$, ${b}_{\scriptscriptstyle a,m}^{*}$, ${x}_{\scriptscriptstyle k,a,m}^{*}$, ${\gamma}_{\scriptscriptstyle k,a,m}^{*}$, ${z}_{\scriptscriptstyle k,a,m}^{*}$
  \\ \textit{Init:}$\bar{\boldsymbol{w}}_{\scriptscriptstyle k,a,m}$, $\bar{{b}}_{\scriptscriptstyle a,m}$, $\bar{{x}}_{\scriptscriptstyle k,a,m}$, $\bar{{\gamma}}_{\scriptscriptstyle k,a,m}$, $\bar{{z}}_{\scriptscriptstyle k,a,m}$, $i=1$, $I_{max}$, $\epsilon$, $err = 1$ 
  \WHILE {$err > \epsilon$ and $i<I_{max}$}
  \STATE Solve \eqref{eq19a} to get $\boldsymbol{w}_{k,a,m}^*$, ${b}_{a,m}^*$, ${x}_{k,a,m}^*$, ${\gamma}_{k,a,m}^*$, ${z}_{k,a,m}^*$\STATE Compute $\displaystyle t^{(i)}$
  \STATE Compute $err = |{t^{(i)}-t^{(i-1)}}|$
  \STATE Update $\bar{\boldsymbol{w}}_{k,a,m} \leftarrow {\boldsymbol{w}}_{\scriptscriptstyle k,a,m}^{*}$; $\bar{{b}}_{\scriptscriptstyle a,m} \leftarrow {{b}}_{\scriptscriptstyle a,m}^{*}$; $\bar{{x}}_{\scriptscriptstyle k,a,m} \leftarrow {{x}}_{\scriptscriptstyle k,a,m}^{*}$; $\bar{\gamma}_{\scriptscriptstyle k,a,m} \leftarrow {\gamma}_{\scriptscriptstyle k,a,m}^*$; $t^{\scriptscriptstyle (i-1)} \leftarrow t^{\scriptscriptstyle (i)}$; $i\leftarrow i+1$ 
  \ENDWHILE
  \end{algorithmic}
  \end{algorithm}
  \end{minipage}
  \end{figure}

 \section{Handover Scenario and Channel Prediction}
 \label{sec:HO}

\begin{figure*}[!t]
  \begin{minipage}{.37\textwidth}
    \centering
    \includegraphics[width=\linewidth, height= 6cm]{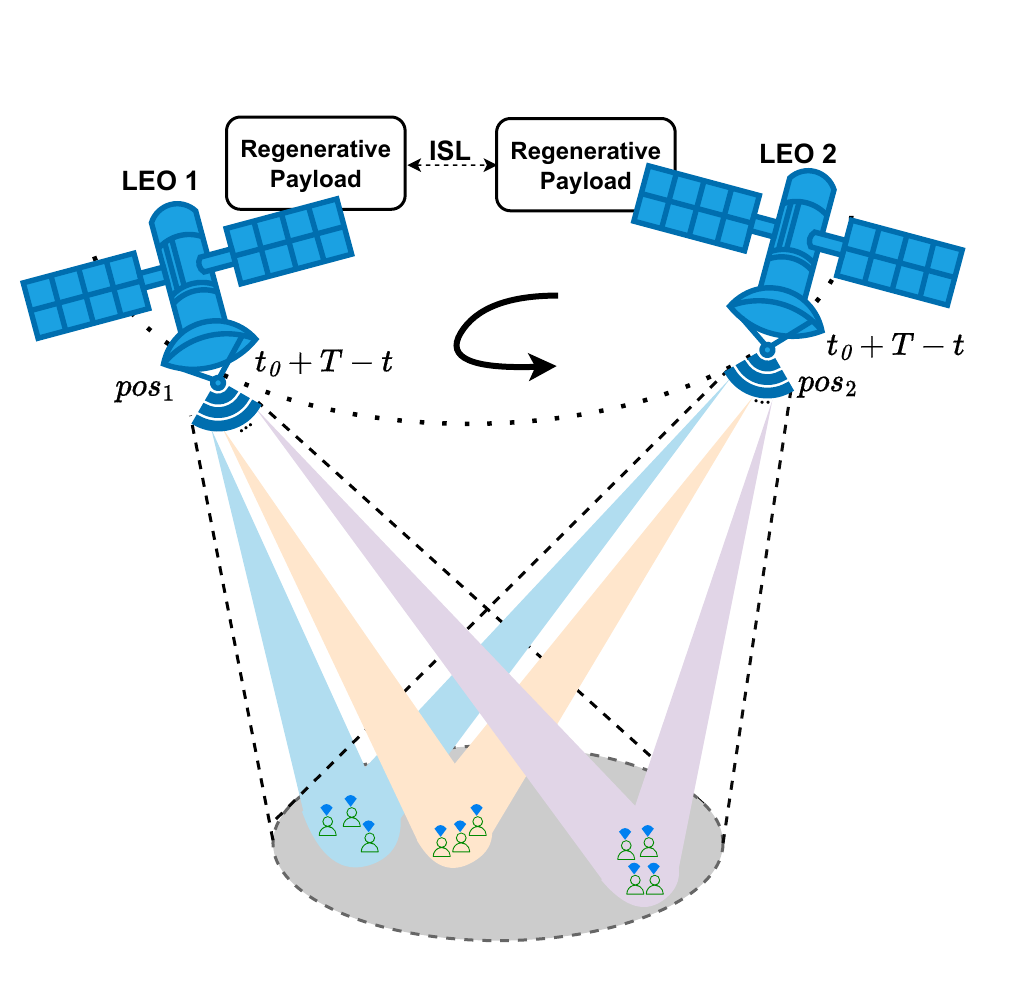}
    \caption{Handover scenario in a downlink multicasting LEO satellite system.}
    \label{fig2}
  \end{minipage}%
  \begin{minipage}{.63\textwidth}
    \centering
    \includegraphics[width=\linewidth, height = 6cm]{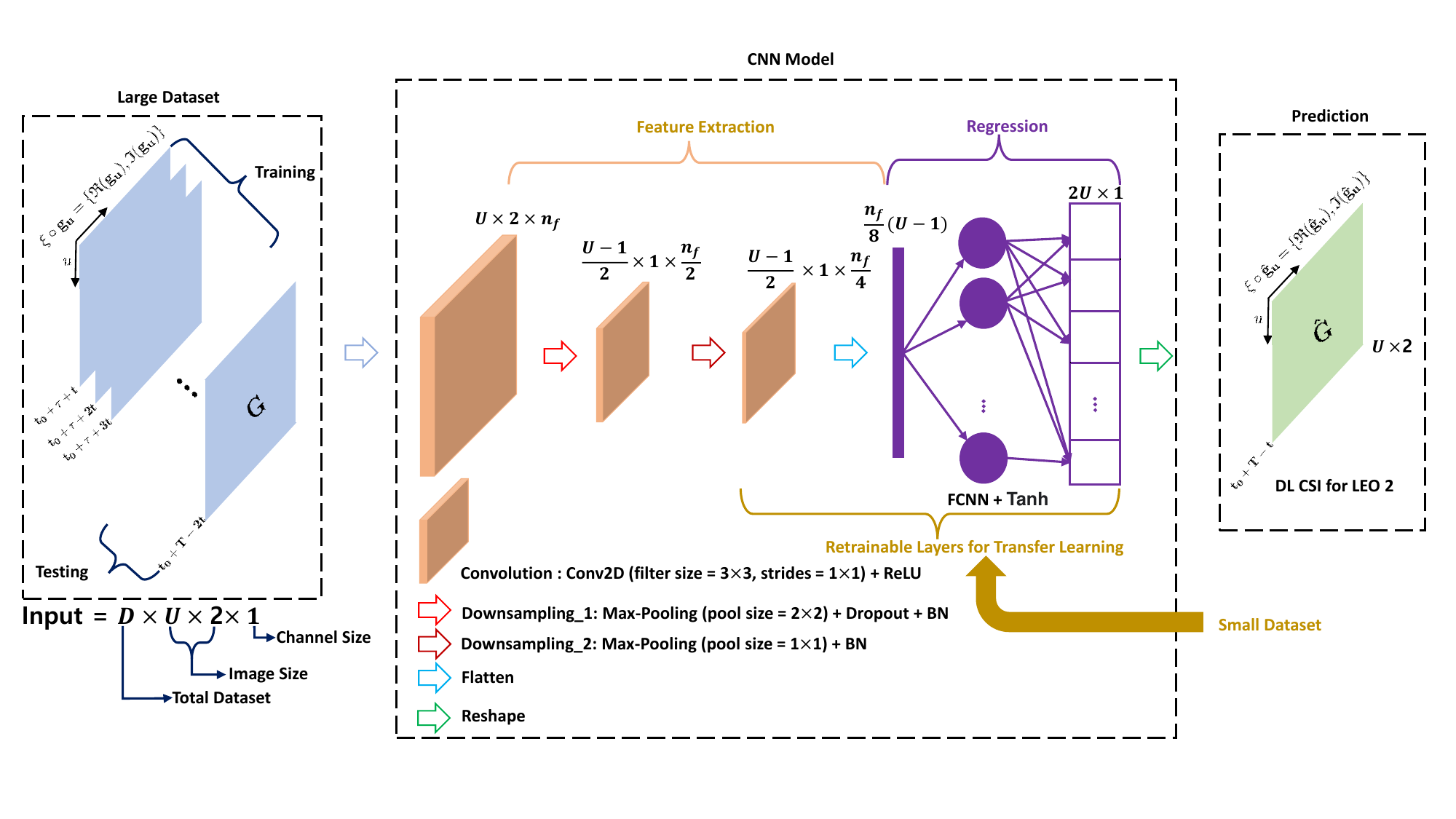}
    \caption{DL-based 2D-CNN model for downlink CSI prediction.}
    \label{fig3}
  \end{minipage}%
\end{figure*}

Due to the short service duration of each LEO satellite pass, HO is important in LEO satellite constellations to guarantee a smooth service. 
% It is a seamless process of transferring a connected data session or mobile call from one LEO satellite to another without interruption, ensuring uninterrupted service to the connected UEs. 
To ensure proper HO in LEO satellite networks, the satellites involved in the HO process must have sufficient time to communicate with each other via an ISL, provided that they are capable of providing service to the same target area \cite{b17}. Additionally, at the beginning of the HO period, the first (departing) LEO satellite informs the second (approaching) LEO satellite about the allocation of spot beams.

\subsection{Joint Transmission during the Handover Period}
To improve the service performance during the HO period, we propose a joint transmission scheme, in which two satellites are jointly sending data to the same UEs for spatial diversity, assuming that the LEOs involved in the HO process are perfectly synchronized, as depicted in Fig.~\ref{fig2}. 
% As shown in the figure, two LEO satellites, one at the end of its service duration and the other entering its service time, are jointly sending data to the same UEs for spatial diversity, assuming that the LEOs involved in the HO process are perfectly synchronized. 
Denote $\mathtt{X} \triangleq \{k,a,m\}$, the signal received by UE $u$ can be written as: 
 \begin{align}
\begin{split}
\small
y_{u,\mathtt{X}}^{HO}=(\boldsymbol{h}_{1,u,\mathtt{X}}^{H}\boldsymbol{w}_{1,\mathtt{X}}+\boldsymbol{h}_{2,u,\mathtt{X}}^{H}\boldsymbol{w}_{2,\mathtt{X}}){s}_{\mathtt{X}}+\check{I}_{m}^{HO}+\hat{I}_{m}^{HO}+ n_{u},   
\end{split}
\end{align}
where $\boldsymbol{h}_{i,u,\mathtt{X}}$ and $\boldsymbol{w}_{i,\mathtt{X}}$ are the downlink channel coefficient and precoding vectors from LEO satellite $i = 1,2$ to the target UE, and $\check{I}_{m}^{HO}$ and $\hat{I}_{m}^{HO}$ are the intra-spot beam interference and inter-spot beam interference, respectively.
$\check{I}_{m}^{HO}$ and $\hat{I}_{m}^{HO}$ are defined as follows:
\begin{align}
\begin{split}
\small
\check{I}_{m}^{HO} \triangleq & {\sum}_{k'\in\mathcal{K}_{a,m} \symbol{92}k} \left( \boldsymbol{h}_{1,u,\mathtt{X}}^{H} \boldsymbol{w}_{1,\mathtt{X}'} + \boldsymbol{h}_{2,u,\mathtt{X}}^{H} \boldsymbol{w}_{2,\mathtt{X}'} \right)s_{\mathtt{X}'},
\end{split} \\
\hat{I}_{m}^{HO} \triangleq &{} \left(\eta_{1,m} P_{1,\sum_{m'}}/B +  \eta_{2,m} P_{2,\sum_{m'}}/B+N_{0} \right) b_{a,m}^{HO}, 
\end{align}
where $\mathtt{X}' \triangleq \{k',a,m\}$ , $\eta_{1,m}$ and $\eta_{2,m}$ represent the aggregated inter-spot beam attenuation factor and the free-space path loss radiated by LEO satellite 1 and 2, respectively. Similarly, $({P_{1,\sum_{m'}}}/{B})$ and $({P_{2,\sum_{m'}}}/{B})$ represent the accumulated interference density caused by the adjacent spot beams radiated by LEO satellite's 1 and 2, respectively, and $b_{a,m}^{HO}$ is the bandwidth of AUG $a$ within spot beam $m$ during the HO period.

The SINR during the HO period can be written as $\gamma_{u,\mathtt{X}}^{HO}=$
\begin{equation}\label{eq23}
\small
\begin{split}
 {{\dfrac{|\boldsymbol{h}_{1,u,\mathtt{X}}^{H}\boldsymbol{w}_{2,\mathtt{X}}+\boldsymbol{h}_{2,u,\mathtt{X}}^{H}\boldsymbol{w}_{2,\mathtt{X}}|^2}{\sum_{k'\in\mathcal{K}_{a,m} \symbol{92}\{k\}}|{\boldsymbol{h}_{1,u,\mathtt{X}}^{H}\boldsymbol{w}_{1,\mathtt{X}'}}+{\boldsymbol{h}_{2,u,\mathtt{X}}^{H}\boldsymbol{w}_{2,\mathtt{X}'}}|^2+\hat{I}_{agg}^{HO}b_{a,m}^{HO}}}},
\end{split}
\end{equation}
where $\hat{I}_{agg}^{HO} \triangleq \left(\eta_{1,m} ({P_{1,\sum_{m'}}}/{B}) +  \eta_{2,m} ({P_{2,\sum_{m'}}}/{B} )+N_{0} \right)$.

The minimum effective transmission rate during the HO period can be expressed as:
 \begin{equation}\label{eq24}
{R_{\mathtt{X}}^{HO}}= b_{a,m}^{HO}\Phi\log_{2}\left(1+ \min_{u}\left\{\gamma_{u,\mathtt{X}}^{HO} \right\}\right).
\end{equation}

It is worth noting that the achievable rate in \eqref{eq24} can be only realized if the precoding vectors are properly designed, which requires the CSI from both LEO satellites. Due to the difference in operating frequencies between the uplink and downlink in satellite communications, the UEs must provide feedback on the downlink CSI to the LEO satellite to design the precoding vectors. 
% In particular, the LEO satellite sends pilot sequences on the downlink to the UEs. 
% Upon the CSI measurements, the UEs feedback the CSI to the LEO satellite on the uplink frequency bandwidth. 

During the HO process, LEO satellite 1, located at position $pos_{1}$, sends pilot signals to single-antenna UEs to estimate downlink CSI and maintain active links. The UEs provide the estimated CSI to LEO satellite 1, which then applies precoding and initiates data transmission. The estimated downlink CSI is assumed to be perfect and remains unchanged when received by LEO satellite 1. Meanwhile, LEO satellite 2 at position $pos_{2}$ uses a DL-based model to determine the downlink channel, which allows synchronized transmission of the same data symbols as LEO satellite 1. In Section V, we explore various HO techniques to achieve this synchronization.

\subsection{DL-based Downlink CSI Prediction}
The conventional communication protocol is not designed to facilitate joint transmission between two LEO satellites, where the channel estimation period is designated to estimate CSI from one LEO satellite at a time. With a single antenna, the UEs cannot estimate the CSI from both LEO satellites without having the communication protocol modified, e.g., a change in the frame structures. To avoid such modification and minimize the CSI estimation time, we propose a DL-based channel prediction scheme applied to LEO 2 (the departing satellite) during the HO period, and apply channel estimation to LEO 1 (the entering satellite), as shown in Fig.~\ref{fig3}. Since the departing satellite has already served the UEs in the current serving period, we can utilize the historical CSI estimates to predict the CSI during the HO time. On the other hand, the entering satellite does not possess any historic CSI measurements. Hence, its CSI can only be estimated through conventional pilot-assisted CSI estimation.

In particular, a DL-based 2D-CNN model \cite{b27} is employed to predict the downlink CSI for LEO satellite 2 at position $pos_2$. The downlink CSI, i.e., ${\boldsymbol{h}}_{u}$ depends on the downlink channel gain, i.e., ${g}_{u}$ and array response vector, i.e., $\boldsymbol{v}_{u}$$(\varphi_u)$. Taking into account that the UEs position is static, $\boldsymbol{v}_{u}$$(\varphi_u)$ can be pre-determined based on the position of the LEO satellite and the position of the UE. Thus, to predict $\hat{g}_{u}$ of $U$ UEs for HO duration, i.e., $(t_{\circ}+T-\tau)$ at once, the ${g}_{u}$ of $U$ UEs during time $t$ are vertically stacked to form ${\boldsymbol{G}}_{t}$ $\in \mathbb{C}^{U \times 1}$ matrix. The historical data of the $\boldsymbol{G}$ is taken as image input data with two channels, then processed by the $l_c$ convolution layers sequentially, which is three in our case, then flattened and processed by the single fully connected neural network (FCNN), and finally reshaped to get $\boldsymbol{\hat{G}}_{t_{\circ}+T-\tau}$. The $u$-{th} row of $\boldsymbol{\hat{G}}_{t_{\circ}+T-\tau}$ corresponds to $\hat{g}_{u,t_{\circ}+T-\tau}$. Thus, $\boldsymbol{\hat{h}}_{u,t_{\circ}+T-\tau}$ $= \boldsymbol{\hat{G}}_{t_{\circ}+T-\tau}(u,:) \cdot$ $\boldsymbol{v}_{u,t_{\circ}+T-\tau}$. To meet the operational requirements of the neural network, we introduce the operator $\xi$ to map the $\boldsymbol{{G}}$ from the complex domain to the real domain, i.e., $\xi \circ \boldsymbol{{G}} = \{\mathfrak{R}(\boldsymbol{G}),\mathfrak{I}(\boldsymbol{G})\}$. The real part $\mathfrak{R}(\boldsymbol{{G}})$ and imaginary part $\mathfrak{I}(\boldsymbol{{G}})$  can be considered as the first channel and second channel, respectively. In addition, the inverse mapping of the operator $\xi$ is $\xi^{-1}$.

In Fig.~\ref{fig3}, the CNN-based downlink CSI (CNN-CSI) prediction model utilizes convolutional layers to extract spatial features from the channel gain. The number of filters in the convolutional layers is set to $n_f$, ${n_f}/{2}$, and ${n_f}/{4}$ respectively. The first convolutional layer takes a 2D input of size $U \times 2$ and uses $n_f$ filters of size $3 \times 3$ with a stride of $1 \times 1$. The output is then passed through a downsampling layer using max-pooling with a pool size of $2 \times 2$, along with batch normalization (BN) and dropout techniques \cite{b34},\cite{b35}. The resulting downsampled data is fed into the second convolutional layer, which further reduces the number of filters. The process is repeated in the third convolutional layer, resulting in feature maps of size $({(U-1) \times 1 \times n_f)}/{(2 \times 4)}$. These features are flattened and passed to a fully-connected neural network (FCNN) for regression. Each convolutional layer utilizes a rectified linear unit (ReLU) activation function to introduce non-linearity. The FCNN layer uses the hyperbolic tangent (tanh) activation function to produce outputs in the range of [-1, 1] \cite{b36}. Finally, the results of the FCNN layer are reshaped into the real and imaginary part of the channel gain matrix to get the predicted output $\{\mathfrak{R}(\hat{\boldsymbol{G}}_{t_{\circ}+T-\tau}),\mathfrak{I}(\hat{\boldsymbol{G}}_{t_{\circ}+T-\tau})\}$ of size $U \times 2$.  Then, the future downlink CSI can be obtained, i.e., $\boldsymbol{\hat{h}}_u (t_{\circ}+T-\tau) = (\xi^{-1} \circ \{\mathfrak{R}(\hat{\boldsymbol{G}}_{t_{\circ}+T-\tau}(u,:),\mathfrak{I}(\hat{\boldsymbol{G}}_{t_{\circ}+T-\tau}(u,:))\}) \cdot \boldsymbol{v}_{u}(t_{\circ}+T-\tau)$.

Our regression problem utilizes the mean square error (MSE) \cite{b37} method for training, and the Adam optimizer \cite{b38} is employed for weight and learning rate updates. The training of the CNN-CSI model follows the mini-batch gradient descent approach, where the dataset of size $\mathpzc{D}$ is divided into $\mathpzc{D}$ batches of size one. Consequently, the loss function based on the data points (pixel)-based MSE \cite{b39}, for each batch $\mathpzc{d}$ ($\mathpzc{d} \in \mathpzc{D}$) is computed as follows:
\begin{align}\label{eq25}
{\mathcal{L}_\mathpzc{d}(\Theta)\!=\! \dfrac{{\|\mathfrak{R}(\boldsymbol{G}_\mathpzc{d})\! -\!\mathfrak{R}(\hat{\boldsymbol{G}}_\mathpzc{d}})\|^{2}\! +\! {\| \mathfrak{I}(\boldsymbol{G}_\mathpzc{d})\! -\!\mathfrak{I}(\hat{\boldsymbol{G}}_\mathpzc{d})\|^{2}}}{{d\times U\times 2}},}
\end{align}
where $d$ in the denominator represents the mini-batch size, and $U \times 2$ represents the total number of data points that make up $\boldsymbol{G}$ (including both real and imaginary parts).

The CNN-CSI model weights ($\Theta$) are updated after each batch by minimizing the loss function $\mathcal{L}_{\mathpzc{d}}(\Theta)$. To reduce training time and propagation delay in the live network, transfer learning is employed. Initially, the CNN-CSI model is trained at the GW. Then, a new CNN model is created by freezing the first two layers of the previous model (Fig.~\ref{fig3}). This new model is trained again on the computationally constrained LEO satellite 2, transferring more general features learned by the initial layers just before the HO period.

\subsubsection{Complexity of the Proposed Algorithm}
The computational complexity to train the CNN-CSI model with $l_c$ layers is given by $\mathcal{O}(E_{max}\mathpzc{D}\sum_{l= 1}^{l_c}n_{f,{l-1}}s_{f,l}^2n_{f,l}2U)$ \cite{b40}, where $n_{f,l}$ 
is the number of filters in the $l$-th layer, $s_{f,l}$ is the spatial size of the 
filters in the $l$-th layer, $\mathpzc{D}$ is the total number of batches, and $E_{max}$ is the maximum number of training epochs required to train the model.

\section{Joint Precoding Vectors Design during the Handover Period}
In this section, we present the joint design of the precoding vectors at the two LEO satellites during the HO time, given the predicted CSI from the previous section. Perfect synchronization is assumed between the two LEO satellites during the HO. We propose two collaboration schemes for computing the precoding vectors: i) \emph{centralized collaboration (CC)}, in which the precoding vectors are computed centrally \cite{b21},\cite{b41},\cite{b42} at the GW and ii) \emph{distributed collaboration (DC)}, in which two LEO satellites jointly compute the precoding vectors via ISL link without using the GW.
% During the HO period, synchronization of data packets between LEOs is necessary. This synchronization can be achieved through collaboration between SNs. Based on the collaboration schemes, the HO is therefore divided into two categories, i.e., distributed collaboration-based HO (DC-HO) and full collaboration-based HO (FC-HO). 

%
% #####################
\subsection{Centralized Collaboration}
\label{sec:HO1}
In this collaboration mode, all the computation is performed centrally at the GW which requires CSI feedback from the LEO satellites. From the system point of view, two LEO satellites are considered as parts of the \emph{compound} antenna arrays of size $2N$. Denote $\boldsymbol{h}_{jnt,u,k,a,m} = [\boldsymbol{h}_{1,u,k,a,m}^{H}, \boldsymbol{\hat{h}}_{2,u,k,a,m}^{H}]^H \in \mathbb{C}^{2N\times1}$  as the aggregated channel gains from two LEO satellites to the UE. We aim to design the optimal precoding vector $\boldsymbol{w}_{jnt,k,a,m} \in$ $\mathbb{C}^{2N\times1}$ for user group $k$ applied to both LEO satellites.

The effective achievable rate of group $k$ of associated group $a$ of spot beam $m$ during the CC-based HO (CC-HO) period, using optimal-based precoding design can be given as:
\begin{align}
\small
R_{k,a,m}^{CC, opt}
={}& \,  b_{a,m}^{opt} \Phi \log_2 (1+ \min_u\{\gamma_{u,k,a,m}^{CC,opt}\}  ) 
\end{align}
where \resizebox{.89\hsize}{!}{$\gamma_{u,k,a,m}^{CC,opt} \triangleq { \frac{|\boldsymbol{h}_{jnt,u,k,a,m}^{H}\boldsymbol{w}_{jnt,k,a,m}|^2}{ {\sum_{k'\in\mathcal{K}_{a,m} \symbol{92}\{k\}}|{\boldsymbol{h}_{jnt,u,k,a,m}^{H}\boldsymbol{w}_{jnt,k',a,m}}|^2+{\hat{I}_{agg}^{HO}b_{a,m}^{opt}}}} }$}, wherein $b_{a,m}^{opt}$ and $\boldsymbol{w}_{jnt,k,a,m}$ are the bandwidth allocation and joint precoding vectors, respectively. 

Although each LEO satellite is seen as parts of the compound antenna array of dimension $2N\times 1$, there are specific restrictions in designing the precoding vectors  $\boldsymbol{w}_{jnt,k,a,m}$ in meeting the per-LEO satellite transmit power constraints. Because the first $N$ rows of $\boldsymbol{w}_{jnt,k,a,m}$ will be applied to the LEO satellite 1 and the last $N$ rows are applied to the LEO satellite 2, we introduce binary diagonal selection matrices $\boldsymbol{J}_1 = \text{diag}([\boldsymbol{1}_{N},\boldsymbol{0}_{N}]) \in \{0,1\}^{2N \times 2N}$ and $\boldsymbol{J}_2 = \text{diag}([\boldsymbol{0}_{N},\boldsymbol{1}_{N}]) \in \{0,1\}^{2N \times 2N}$.  Then, the joint bandwidth and precoding vectors design can be formulated as follows:
\begin{subequations}\label{eq27}
\small
\begin{align}
\mathcal{P}^{CC,opt}: {}& \min_{\substack{\{\boldsymbol{w}, \boldsymbol{b}\}}} ~~ \max_{k,a,m}\left( \max \left(\frac{q_k}{R^{CC,opt}_{\mathtt{X}}}+\frac{D_k}{c}, \Pi_{k} \right) \right) \label{eq27a} \\
\text{s.t.} {}& \, R_{\mathtt{X}}^{CC,opt} \ge R_{req}, \forall k,a,m,  \label{eq27b} \\
{}&\,  {{ \sideset{}{_{k \in \mathcal{K}_{a,m}}}\sum{ \sideset{}{_{a \in \mathcal{A}_{m}}}\sum \lVert \boldsymbol{J}_1\boldsymbol{w}_{jnt, \mathtt{X}}  \rVert^{2} \le \frac{P_{\sum}K_{m}}{K}}},  \forall m, \label{eq27c} }\\
{}& {{ \sideset{}{_{k \in \mathcal{K}_{a,m}}}\sum{ \sideset{}{_{a \in \mathcal{A}_{m}}}\sum \lVert \boldsymbol{J}_2\boldsymbol{w}_{jnt,\mathtt{X}}  \rVert^{2} \le \frac{P_{\sum}K_{m}}{K}}},  \forall m, \label{eq27d} }\\
{}& \, {\sideset{}{_{a\in \mathcal{A}_{m}}}\sum b_{a,m}^{opt}} \le B, \quad \forall m, \label{eq27e}
\end{align}
\end{subequations}
where $\mathtt{X} \triangleq \{k,a,m\}$, $\boldsymbol{w} \triangleq \{\boldsymbol{w}_{jnt,\mathtt{X}}\}_{\forall k,a,m}$, and $\boldsymbol{b} \triangleq \{b_{a,m}^{opt}\}_{\forall a,m}$ are the short-hand notations for indexes, precoding vectors and bandwidth allocation, respectively. 

We observe that problem \eqref{eq27} is similar to problem \eqref{eq11} except constraints \eqref{eq27c} and \eqref{eq27d}. Fortunately, these constraints are convex, thus we can adopt the similar technique developed in Section~\ref{sec:jointframework}. Upon obtaining the optimal precoding vectors  $\boldsymbol{w}_{jnt,k,a,m}$, the GW sends the corresponding precoding coefficients to the two LEO satellites for data transmission.

\subsection{Distributed Collaboration}
\label{sec:HO2}
Although the centralized collaboration scheme offers the optimal precoding vectors, it requires excessive signalling overhead, which motivates us to propose the distributed precoding design scheme. In this scheme, each LEO satellite compute the bandwidth allocation and precoding vectors based only on its local CSI and limited exchanged information from the other LEO satellite. Assuming a high-capacity ISL, the two LEO satellites iteratively improve its solutions via iterations. It is noted that the LEO satellites in the distributed collaboration only exchange their power scaling factors, while the bandwidth resource per user group is optimized locally.

To minimize the exchanged overhead and computation load, we consider ZF-based joint bandwidth and power allocation in this scenario. 
% In the DC-based scheme, for the low-complexity hardware design, ZF-based precoding is applied to determine the precoding vectors of LEOs 1 and 2, when they are at the positions $pos_1$ and $pos_2$, respectively. 
Let $\boldsymbol{W}_{i,a,m} = \boldsymbol{H}_{i,a,m}^{H}(\boldsymbol{H}_{i,a,m}\boldsymbol{H}_{i,a,m}^{H})^{-1}$ denote the ZF-beamforming matrix for AUG $a$ of spot beam $m$ of LEO satellite $i$, $ i = 1, 2$, where $\boldsymbol{H}_{i,a,m}$ is the corresponding aggregated channel matrix. Under the ZF design, the precoding vector designed at LEO satellite $i \in \{1,2\}$ for the group $\mathtt{X} \triangleq \{k,a,m\}$ is given as ${\boldsymbol{w}}_{i,\mathtt{X}}^{ZF} = \sqrt{p_{i,\mathtt{X}}}\tilde{\boldsymbol{w}}_{i,\mathtt{X}}$, where $\tilde{\boldsymbol{w}}_{i,\mathtt{X}}$ is the $k$-th column of the ZF precoding matrix, and $p_{i,\mathtt{X}}$ is the power scaling factor. By definition,  $\boldsymbol{h}_{i,u,k,a,m}^{H}\tilde{\boldsymbol{w}}_{i,k',a,m} = \delta_{k,k'},\forall i,u,a,m$ assuming the accurate CSI estimation and prediction. 
% Since the ${\boldsymbol{\hat{w}}}_{2,k,a,m}^{ZF}$ is designed as per the prediction model, we have  $\boldsymbol{h}_{2,u,k,a,m}^{H}{\boldsymbol{\hat{w}}}_{2,k',a,m} = \delta_{2,u,k,k'}, \forall u,k,k',a,m,$ where $\delta_{2,u,k,k'}$ approaches to 1 when $k = k'$, otherwise approaches to 0, provided that $\boldsymbol{h}_{2,u,k,a,m}^{H} \approx \boldsymbol{\hat{h}}_{2,u,k,a,m}^{H}$. 
As a result, the achievable effective rate during the DC-based HO (DC-HO) period is computed as:
\begin{align}
\small
&R_{\mathtt{X}}^{DC,ZF} \notag\\
% ={}& \,
% b_{a,m}^{DC,ZF} \Phi\log_{2} \left(1+\min_{u}\left\{\gamma_{u,k,a,m}^{DC,ZF}\right\}\right) \nonumber  \\
&= b_{a,m}^{ZF} \Phi \log_2 \Big(1+ \min_u\Big\{{ \left(|\boldsymbol{h}_{1,u,\mathtt{X}}^{H}\boldsymbol{w}_{1,\mathtt{X}}^{ZF}+\boldsymbol{h}_{2,u,\mathtt{X}}^{H}\boldsymbol{{w}}_{2,\mathtt{X}}^{ZF}|^2\right)}/ \notag \\
& \qquad    \Big({\underset{k'\in\mathcal{K}_{a,m} \symbol{92}\{k\}}{\sum}|{\boldsymbol{h}_{1,u,\mathtt{X}}^{H}\boldsymbol{w}_{1,\mathtt{X}'}^{ZF}}+{\boldsymbol{h}_{2,u,\mathtt{X}}^{H}\boldsymbol{{w}}_{2,\mathtt{X}'}^{ZF}}|^2 + \hat{I}_{agg}^{HO}b_{a,m}^{ZF} } \Big) \Big\} \Big)  \nonumber\\
&= ~ b_{a,m}^{ZF} \Phi \log_{2} \Big(1+ \frac{(\sqrt{p_{1,\mathtt{X}}}+\sqrt{p_{2,\mathtt{X}}})^2}{\hat{I}_{agg}^{HO}b_{a,m}^{ZF}} \Big), \forall k,a,m.
\end{align}
%
%where $\mathtt{X}' \triangleq \{k',a,m\}$.
% , $b_{a,m}^{ZF}$ is the bandwidth allocation, and $\boldsymbol{w}_{1,\mathtt{X}}^{ZF}$ and $\boldsymbol{{w}}_{2,\mathtt{X}}^{ZF}$ are the precoding vectors for LEO satellite 1 based on estimation and LEO satellite 2 based on prediction, respectively.

Denote $\alpha_{1,\mathtt{X}} \triangleq \| {\tilde{\boldsymbol{w}}_{1,\mathtt{X}}}\|^{2}$ and $\alpha_{2,\mathtt{X}} \triangleq \| {\boldsymbol{\tilde{w}}}_{2,\mathtt{X}}\|^{2}$, the short-term delivery period minimization during DC-HO period under the ZF design can be formulated as:
\begin{subequations}\label{eq29}
\small
\begin{align}
% \begin{split}
\mathcal{P}^{DC,ZF}: {}&  \min_{\{{\boldsymbol{p}}_{1},{\boldsymbol{p}}_{2},\boldsymbol{b}\}} \max_{_{\{k,a,m\}}} \left( \max \left( \frac{q_k}{R_{\mathtt{X}}^{DC,ZF}}+\frac{D_k}{c}, \Pi_{k} \right)\right) \label{eq29a} \\
% \end{split} \\
\text{s.t.} {}& \,  R_{\mathtt{X}}^{DC,ZF} \ge R_{req}, \quad  \forall k,a,m, \label{eq29b} \\
  {}& \, {{ \sideset{}{_{k \in \mathcal{K}_{a,m}}}\sum{ \sideset{}{_{a \in \mathcal{A}_{m}}}\sum \alpha_{1,\mathtt{X}} p_{1,\mathtt{X}}  \le \frac{P_{\sum}K_{m}}{K}}},  \forall m, \label{eq29c}} \\
 {}& \, {{ \sideset{}{_{k \in \mathcal{K}_{a,m}}}\sum{ \sideset{}{_{a \in \mathcal{A}_{m}}}\sum \alpha_{2,\mathtt{X}} p_{2,\mathtt{X}}  \le \frac{P_{\sum}K_{m}}{K}}},  \forall m, \label{eq29d}}  \\
{}& \, {\sideset{}{_{a\in \mathcal{A}_{m}}}\sum b_{a,m}^{ZF}} \le B, \quad \forall m, \label{eq29e} 
\end{align}
\end{subequations}
where ${\boldsymbol{p}}_{1} \triangleq \{{p_{1,\mathtt{X}}}\}_{\forall k,a,m}$, ${\boldsymbol{p}}_{2} \triangleq \{{p_{2,\mathtt{X}}}\}_{\forall k,a,m}$ and ${\boldsymbol{b}}$ are the short-hand notations. 

The problem $\mathcal{P}^{DC,ZF}$ is non-convex due to the objective function $\eqref{eq29a}$ and $\eqref{eq29b}$, respectively. From the implementation perspective, the computation of the precoding vectors, as well as the bandwidth allocations, have to be executed at each LEO satellite separately. Furthermore, the bandwidth allocation $\boldsymbol{b}$ must be synchronized such that they allocate the same bandwidth to the requesting UEs. To achieve this goal, we propose an iterative algorithm in which two LEO satellites consecutively optimize their power factors and the bandwidth allocation, assuming the output of the other LEO satellite is shared. In the initialization, the problem $\mathcal{P}^{DC,ZF}$ is solved in the LEO 1 considering arbitrary feasible power value $\sqrt{\bar{p}_{2,k,a,m}}$ at LEO 2. The resulting joint optimization problem can be written as:
\begin{subequations}\label{eq30}
\small
\begin{align}
&\mathcal{P}_{LEO1}^{DC,ZF}(\boldsymbol{\bar{p}}_2): {} \min_{\{{\boldsymbol{p}_1},\boldsymbol{b}_1\}} \max_{\{k,a,m\}}\! \Big(\! \max\! \Big(\frac{q_k}{R_{1,\mathtt{X}}^{DC,ZF}}\! +\! \frac{D_k}{c}, \Pi_{k} \!\Big)\!\Big)\label{eq30a} \\
& \text{s.t.}~~  b_{1,a,m} \Phi \log_{2}\left(1 + (\sqrt{p_{1,\mathtt{X}}} + \sqrt{\bar{p}_{2,\mathtt{X}}})^2/(\hat{I}_{agg}^{HO}b_{1,a,m})\right) \nonumber \\ 
& \qquad ~~ \ge R_{req}, \,  \forall k,a,m; \ \eqref{eq29c}, \eqref{eq29e}, \label{eq30b} 
\end{align}
\end{subequations}
where 
% $\mathtt{X} \triangleq \{k,a,m\}$ is the short-hand indexes and 
$R_{1,\mathtt{X}}^{DC,ZF} \triangleq b_{1,a,m}\Phi\log_2 \big(1 + \frac{(\sqrt{p_{1,\mathtt{X}}} + \sqrt{\bar{p}_{2,\mathtt{X}}})^2}  {(\hat{I}_{agg}^{HO} b_{1,a,m})}\big)$.

The main challenge in solving problem $\mathcal{P}_{LEO1}^{DC,ZF}$ lies in the objective function and the first constraint $\eqref{eq30b}$. We can handle the constraint $\eqref{eq30b}$ by considering the slack variables ${x}_{1,k,a,m}$, which can be reformulated as:
\begin{gather}
\small
{b_{1,a,m} \Phi \log_{2}\Big(1+\frac{{x}_{1,k,a,m}}{\hat{I}_{agg} b_{1,a,m}}\Big) \ge R_{req}, \,  \forall k,a,m,\label{eq31}} \\
\sqrt{{p}_{1,k,a,m}} + \sqrt{\bar{p}_{2,k,a,m}} \ge \sqrt{{x}_{1,k,a,m}}. \label{eq32}  
\end{gather}

\begin{proposition}\label{prop1}
The rate function under the ZF design in $\eqref{eq31}$ is jointly concave in ${b}_{1,a,m}$ and ${x}_{1,k,a,m}$.
\end{proposition}
The proof of Proposition 1 is shown in Appendix A. The constraint $\eqref{eq32}$ is a DC form as both sides are convex functions. Thus, it can be efficiently solved using the iterative-based SCA method by taking the first-order approximation of the RHS of the constraint $\eqref{eq32}$. Let $\bar{x}_{1,k,a,m}$ be a feasible value of the constraint $\eqref{eq32}$ in the current iteration. In the next iteration, the constraint $\eqref{eq32}$ can be approximated as a convex constraint as:
\begin{equation}\label{eq33} 
\sqrt{{p}_{1,k,a,m}} + \sqrt{\bar{p}_{2,k,a,m}}  \ge \frac{{x}_{1,k,a,m}}{2\sqrt{\bar{x}_{1,k,a,m}}} + \frac{\sqrt{{\bar{x}}_{1,k,a,m}}}{2}.  
\end{equation}

Now, the problem $\mathcal{P}_{LEO1}^{DC,ZF}$ can be approximated by a convex optimization problem $\mathcal{P}_{LEO1}^{DC,ZF'}(\boldsymbol{{\bar{p}}}_{2},\boldsymbol{{\bar{x}}}_{1})$:
\begin{subequations}\label{eq34}
\begin{align}
\min_{\{ \boldsymbol{p}_1,\boldsymbol{b}_1,\boldsymbol{x}_1\}} &~ \max_{\{k,a,m\}} \big( \max \big( \frac{q_k}{{R_{1,\mathtt{X}}^{DC,ZF}}^{'}}+\frac{D_k}{c}, \Pi_{k} \big)\big)\label{eq34a}\\
\text{s.t.} & ~~ \eqref{eq31},\eqref{eq33}, \eqref{eq29c}, \eqref{eq29e}, \nonumber
\end{align}
\end{subequations}
where ${\boldsymbol{{x}}_{1}} \triangleq \{{{x}_{1,k,a,m}}\}_{\forall k,a,m}$ and ${\boldsymbol{\bar{x}}_{1}} \triangleq \{{\bar{x}_{1,k,a,m}}\}_{\forall k,a,m}$ are the short-hand notations.
% for the slack variable and initialization of the slack variable, respectively, and ${R_{1,\mathtt{X}}^{DC,ZF}}^{'} = {b_{1,a,m} \Phi \log_2\left(1+x_{1,\mathtt{X}} / (\hat{I}_{agg}^{HO} b_{1,a,m})\right)}$.

Let ${p}_{1,\mathtt{X}}^{*}$ be the solution of problem \eqref{eq34}. This value will be communicated to the LEO 2 to compute its optimal transmit power, i.e., ${p}_{2,\mathtt{X}}$. It is worth noting that there is no need to exchange the bandwidth allocation $b_{1,a,m}$ to perform the optimization in LEO satellite 2, but only for checking the termination criteria. 
% between the two LEOs, as we can show later via numerical results that both LEOs will converge to the same optimal, i.e., $b_{a,m}^\star$.
The optimization problem in LEO 2 can be formulated similarly as in \eqref{eq30} with the satellite subscript $1$ switched with $2$. 
%
% \begin{subequations}\label{eq35}
% \small
% \begin{align}
% \begin{split}
% \small
% \mathcal{P}_{LEO2}^{DC,ZF}({\boldsymbol{\bar{p}_}_1}): {}& \min_{\{{\boldsymbol{p}_2},\boldsymbol{b}_2\}}  \max_{\{k,a,m\}} \Bigg( \max \left(\frac{q_k}{R_{2,\mathtt{X}}^{DC,ZF}} + \frac{D_k}{c}, \Pi_{k} \right)\Bigg)\label{eq35a}
%  \end{split}\\
% \text{s.t.} {}& \, b_{2,a,m} \Phi \log_{2}\left(1 + (\sqrt{\bar{p}_{1,\mathtt{X}}} + \sqrt{{p}_{2,\mathtt{X}}})^2/(\hat{I}_{agg}^{HO}b_{2,a,m})\right) \nonumber \\
% {}& \, \ge R_{req}, \,  \forall k,a,m, \label{eq35b} \\
% {}& \, \eqref{eq29d}, \eqref{eq29e} \nonumber
% %
% %
% % \begin{split}
% % \mathcal{P}_{LEO2}^{DC,ZF}({\boldsymbol{\bar{p}}_{1}}): {}& \min_{_{\{{\boldsymbol{\tilde{p}}_{2}},\boldsymbol{\tilde{b}}_{2}  \}}}  \quad \max_{_{\{k,a,m\}}} \left( \max \left( \frac{q_k}{\boldsymbol{\tilde{b}}_{2} \Phi \log_{2}\left(1+{(\sqrt{{\boldsymbol{\bar{p}}_{1}}}+\sqrt{{\boldsymbol{\tilde{p}}_{2}}})^2}/({\hat{I}_{agg}^{HO}\boldsymbol{\tilde{b}}_{2}})\right)}+\frac{D_k}{c}, \Pi_{k} \right)\right)\label{eq33a}
% % \end{split} \\
% % \text{s.t.} {}& \, \boldsymbol{\tilde{b}}_{2} \Phi \log_{2}\left(1+{(\sqrt{{\boldsymbol{\bar{p}}_{1}}}+\sqrt{{\boldsymbol{\tilde{p}}_{2}}})^2}/({\hat{I}_{agg}^{HO}\boldsymbol{\tilde{b}}_{2}})\right) \ge R_{req}, \,  \forall k,a,m, \label{eq33b} \\
% % {}&\eqref{eq27d}, \eqref{eq27e}, \nonumber
% \end{align}
% \end{subequations}
% %
% where
% $R_{2,\mathtt{X}}^{DC,ZF} \triangleq b_{2,a,m} \Phi \log_2 \left(1 + \frac{(\sqrt{\bar{p}_{1,\mathtt{X}}} + \sqrt{{p}_{2,\mathtt{X}}})^2}  {(\hat{I}_{agg}^{HO} b_{2,a,m})}\right)$.
%
Following the same method, the problem $\mathcal{P}_{LEO2}^{DC,ZF}$ can be solved by using the SCA approach of its approximated convex problem $\mathcal{P}_{LEO2}^{DC,ZF'}({\boldsymbol{\bar{p}}_{1}},{\boldsymbol{\bar{x}}_{2}})$:
\begin{subequations}\label{eq36}
\begin{align}
&\min_{\{{\boldsymbol{{p}}_{2}},\boldsymbol{b}_2,\boldsymbol{{x}}_{2}\}} ~ \max_{\{k,a,m\}}\Big( \max \Big( \frac{q_k}{{R_{2,\mathtt{X}}^{DC,ZF}}^{'}} + \frac{D_k}{c},\Pi_{k} \Big)\Big)\label{eq36a} \\
&\text{s.t.} ~
\, b_{2,a,m} \Phi \log_{2}\big(1 + x_{2,\mathtt{X}}/(\hat{I}_{agg}^{HO} b_{2,a,m})\big) 
\ge R_{req}, \,  \forall k,a,m, \notag \\
&~~ \sqrt{\bar{p}_{1,\mathtt{X}}}\! +\! \sqrt{{p}_{2,\mathtt{X}}}  \ge \frac{{x}_{2,\mathtt{X}}}{\sqrt{\bar{x}_{2,\mathtt{X}}}}\! +\! \frac{\sqrt{{\bar{x}}_{2,\mathtt{X}}}}{2}, \forall k,a,m;\ \eqref{eq29d}, \eqref{eq29e}, \nonumber
\end{align}
\end{subequations}
where 
% ${\boldsymbol{{x}}_{2}} \triangleq \{{{x}_{2,k,a,m}}\}_{\forall k,a,m}$ and ${\boldsymbol{\bar{x}}_{2}} \triangleq \{{\bar{x}_{2,k,a,m}}\}_{\forall k,a,m}$ are the short-hand notations for the slack variable and initialization of the slack variable, respectively, and 
$R_{2,\mathtt{X}}^{DC,ZF^{'}} \triangleq \Phi_{a,m} b_{2,a,m} \log_2 \big(1+x_{2,\mathtt{X}} / (\hat{I}_{agg}^{HO} b_{2,a,m}) \big)$.

The optimal output powers $p_{2,\mathtt{x}}^*$ of $\mathcal{P}_{LEO2}^{DC,ZF'}$ will be forwarded to LEO satellite 1 to execute the next iteration of optimization. The iterations will continue until the convergence of the optimal transmit powers and bandwidth allocation. The detailed steps of the proposed algorithms are presented in Algorithms 3 and 4.
% If $\tilde{b}_{2,a,m}^{DC,ZF}$ from $\mathcal{P}_{LEO2}^{DC,ZF'}$ converges to $\tilde{b}_{1,a,m}^{DC,ZF}$ from $\mathcal{P}_{LEO1}^{DC,ZF'}$, then $\mathcal{P}_{LEO2}^{DC,ZF'}$ is the final solution. Otherwise, ${\bar{p}}_{1,k,a,m}$ and ${\bar{p}}_{2,k,a,m}$ is exchanged between the LEOs until $\tilde{b}_{1,a,m}^{DC,ZF} \approx \tilde{b}_{2,a,m}^{DC,ZF}$. The solution from $\mathcal{P}_{LEO2}^{DC,ZF'}$ is sub-optimal for $\mathcal{P}^{DC,ZF}$ and heavily depends on the initialization of ${\boldsymbol{\bar{p}}_{2}}$, ${\boldsymbol{\bar{x}}_{1}}$, and ${\boldsymbol{\bar{x}}_{2}}$. Therefore, Algorithms 3 and 4 are proposed to solve $\eqref{eq27}$.
%
\begin{figure*}[b!]
\small
  \centering
  \begin{minipage}{0.446\textwidth}
    \centering
    \begin{algorithm}[H]
      \caption{Iter. Alg. to Solve \eqref{eq29a}}
      \label{algorithm:3}
      \begin{algorithmic}[1]
     \renewcommand{\algorithmicrequire}{\textbf{Input:}}
     \renewcommand{\algorithmicensure}{\textbf{Output:}}
     \REQUIRE $\mathcal{A}_{m}$, $\mathcal{K}_{a,m}$, $K_m$, $\boldsymbol{h}_{1,u,k,a,m}$, $\boldsymbol{w}_{1,k,a,m}$, $\mu_k$, $D_k$, $c$, $R_{BH}$, $\eta_{1,m}$, ${P_{1,\sum_{m'}}}/{B}$
     \ENSURE ${p}_{\scriptscriptstyle 1,k,a,m}^{*}$, 
     ${b}_{\scriptscriptstyle 1,a,m}^{*}$,
     ${x}_{\scriptscriptstyle 1,k,a,m}^{*}$
     \\ \textit{Initialization:} $\bar{p}_{\scriptscriptstyle 1,k,a,m}$, $\bar{x}_{\scriptscriptstyle 1,k,a,m}$,  ${\bar{b}}_{1,a,m} = 1$, ${\bar{b}}_{2,a,m} = 5$, $i=1$, $I_{max}$, $\epsilon$, $err = 1$ 
     \WHILE {$\displaystyle (\bar{b}_{1,a,m} \mathrel{\mathtt{!=} \bar{b}_{2,a,m}} )$} 
     \STATE Solve $\eqref{eq34}$ at LEO 1 
      \WHILE {$err > \epsilon$ and $i<I_{max}$}
      \STATE Solve \eqref{eq34} to get ${{p}}_{\scriptscriptstyle 1,k,a,m}^{*}$,
     ${{b}}_{\scriptscriptstyle 1,a,m}^{*}$, ${x}_{\scriptscriptstyle 1,k,a,m}^{*}$ 
      \STATE Compute $\displaystyle t^{DC,ZF(i)}_{LEO1}$ \& $err = |{t^{DC,ZF(i)}_{LEO1} - t^{DC,ZF(i-1)}_{LEO1}}|$
    \STATE Update $\bar{{x}}_{\scriptscriptstyle 1,k,a,m} \leftarrow {{x}}_{\scriptscriptstyle 1,k,a,m}^{*}$; $t^{DC,ZF(i-1)}_{LEO1} \leftarrow t^{DC,ZF(i)}_{LEO1}$; $i\leftarrow i+1$ 
    \ENDWHILE
    \STATE Update ${\bar{p}}_{\scriptscriptstyle 1,k,a,m} \leftarrow {{p}}_{\scriptscriptstyle 1,k,a,m}^{*}$, ${\bar{b}}_{\scriptscriptstyle 1,a,m} \leftarrow {{b}}_{\scriptscriptstyle 1,a,m}^{*}$
    \STATE Send/Receive via ISL
    \ENDWHILE    
      \end{algorithmic}
    \end{algorithm}
  \end{minipage}
  \hspace{0.2em} 
  \begin{tikzpicture}[>=stealth]
      \draw[thick,->] (0,-5) -- (1,-5) node[midway, above] {${\bar{p}}_{\scriptscriptstyle {1,k,a,m}}$} node[midway, below] {${\bar{b}}_{\scriptscriptstyle {1,a,m}}$};
      \draw[thick,<-] (0,-6.25) -- (1,-6.25) node[midway, above] {${\bar{p}}_{\scriptscriptstyle {2,k,a,m}}$} node[midway, below] {${\bar{b}}_{\scriptscriptstyle {2,a,m}}$};
  \end{tikzpicture}
  \begin{minipage}{0.45\textwidth}
    \centering
    \begin{algorithm}[H]
      \caption{Iter. Alg. to Solve \eqref{eq29a}}
      \begin{algorithmic}[1]
         \renewcommand{\algorithmicrequire}{\textbf{Input:}}
         \renewcommand{\algorithmicensure}{\textbf{Output:}}
         \REQUIRE $\mathcal{A}_{m}$, $\mathcal{K}_{a,m}$, $K_m$, $\boldsymbol{\hat{h}}_{2,u,k,a,m}$,  $\boldsymbol{\hat{w}}_{2,k,a,m}$, $\mu_k$, $D_k$, $c$, $R_{BH}$, $\eta_{2,m}$, ${P_{2,\sum_{m'}}}/{B}$
         \ENSURE ${p}_{\scriptscriptstyle 2,k,a,m}^{*}$,
         ${b}_{\scriptscriptstyle 2,a,m}^{*}$,
         ${x}_{\scriptscriptstyle 2,k,a,m}^{*}$
         \\ \textit{Initialization:} $\bar{p}_{\scriptscriptstyle 2,k,a,m}$, $\bar{x}_{\scriptscriptstyle 2,k,a,m}$, ${\bar{b}}_{1,a,m} = 1$, ${\bar{b}}_{2,a,m} = 5$, $i=1$, $I_{max}$, $\epsilon$, $err = 1$
         \WHILE {$\displaystyle (\bar{b}_{1,a,m} \mathrel{\mathtt{!=} \bar{b}_{2,a,m}} )$} 
         \STATE Solve $\eqref{eq36}$ at LEO satellite 2 
          \WHILE {$err > \epsilon$ and $i<I_{max}$}
          \STATE Solve \eqref{eq36} to get ${{p}}_{\scriptscriptstyle 2,k,a,m}^{*}$,
         ${{b}}_{\scriptscriptstyle 2,a,m}^{*}$, ${x}_{\scriptscriptstyle 2,k,a,m}^{*}$ 
          \STATE Compute $\displaystyle t^{DC,ZF(i)}_{LEO2}$ \& $err = |{t^{DC,ZF(i)}_{LEO2} - t^{DC,ZF(i-1)}_{LEO2}}|$
        \STATE Update $\bar{{x}}_{\scriptscriptstyle 2,k,a,m} \leftarrow {{x}}_{\scriptscriptstyle 2,k,a,m}^{*}$; $t^{DC,ZF(i-1)}_{LEO2} \leftarrow t^{DC,ZF(i)}_{LEO2}$; $i\leftarrow i+1$ 
        \ENDWHILE
        \STATE Update ${\bar{p}}_{\scriptscriptstyle 2,k,a,m} \leftarrow {{p}}_{\scriptscriptstyle 2,k,a,m}^{*}$, ${\bar{b}}_{\scriptscriptstyle 2,a,m} \leftarrow {{b}}_{\scriptscriptstyle 2,a,m}^{*}$
        \STATE Send/Receive via ISL   
        \ENDWHILE  
      \end{algorithmic}
    \end{algorithm}
  \end{minipage}
\end{figure*}
\subsection{Major Technical Challenges \& Their Solutions}
For the signals emitted by the LEO satellites involved in the HO process introduced in Sections \ref{sec:HO1} and \ref{sec:HO2}, such that the signals add constructively to increase the total signal strength. Some of the major technical challenges that could be encountered when implementing our proposed HO approach in a real system, along with possible directions to address them, are provided below:
\begin{itemize}
\item ISL-based bandwidth synchronization: When two satellites are widely separated, pointing, tracking, and acquisition required to establish ISL connection between them requires onboard special hardware embedded in them \cite{b43}.
% \item \textcolor{blue}{Timing alignment: To achieve precise timing alignment, satellites require oscillators with errors in the range of nanosecond or picosecond range, such as an ultra-stable oscillator. Developing such oscillators can be challenging and expensive.}
\item Difference of slant distance between satellites involved in HO and a reference location \cite{b44}: To address this timing offset before transmission is required, which becomes more challenging when both the transmitter and receiver are in motion.
\item CSI prediction accuracy: In the DC-HO scheme, predicting the CSI for the leaving LEO satellite relies on historical CSI information, the spatial and temporal correlation between data points, and the specific machine learning techniques used for prediction.
% \item \textcolor{blue} {Location accuracy: It is more challenging for the hiding LEO satellite to precisely determine the positions of requesting UEs during the DC-HO period. It relies on the approaching LEO satellite to provide precise UE positions via the ISL link. Achieving this precision requires attention to GPS accuracy in the UEs, user velocity relative to the approaching LEO satellite, and management of signal propagation delays.}
\end{itemize}

%###################################################################
\section{Performance Evaluation on Realistic System Parameters}
In this section, we eveluate the performance of the proposed framework based on realistic LEO satellite parameters and Movielens dataset. 

\subsection{LEO Satellite Footprint}
\begin{figure*}[h]%
    \centering
    \subfloat[\centering]{{\includegraphics[height = 5cm, width = 5cm]{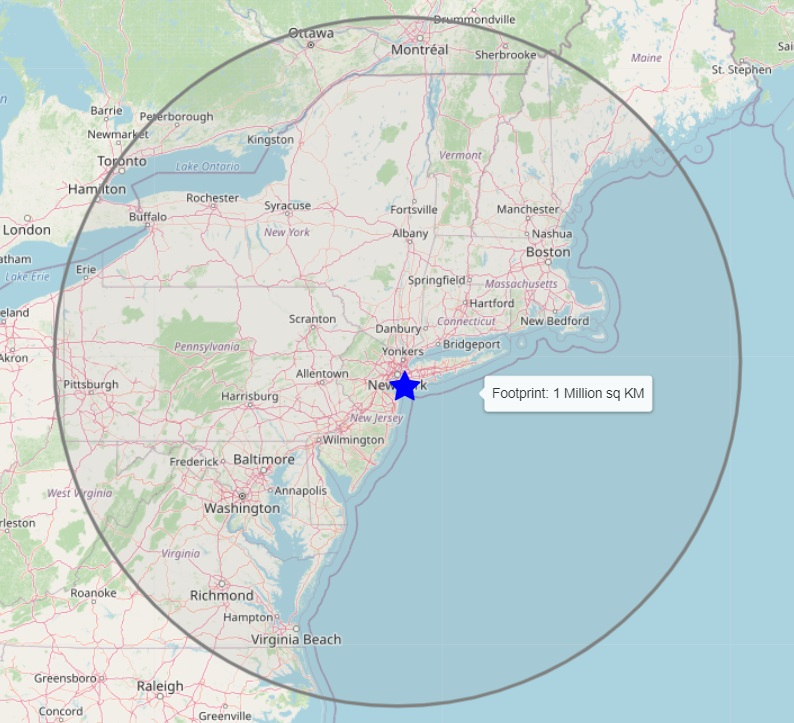} }}%
    \subfloat[\centering]{{\includegraphics[height = 5cm, width = 5cm]{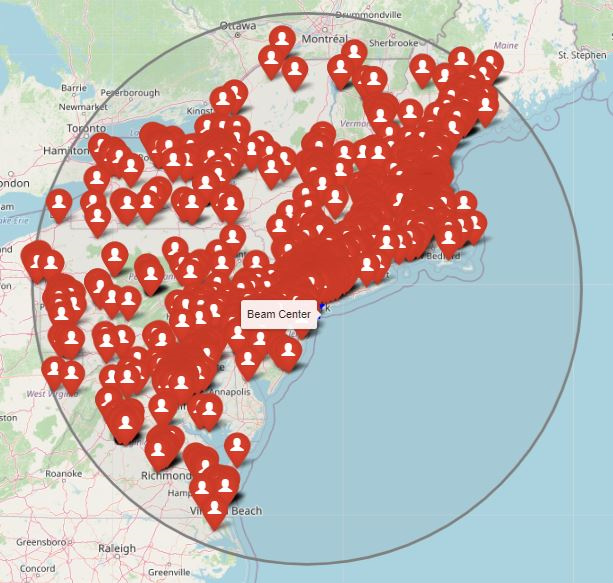} }}%
    \subfloat[\centering]{{\includegraphics[height = 5cm, width = 6cm]{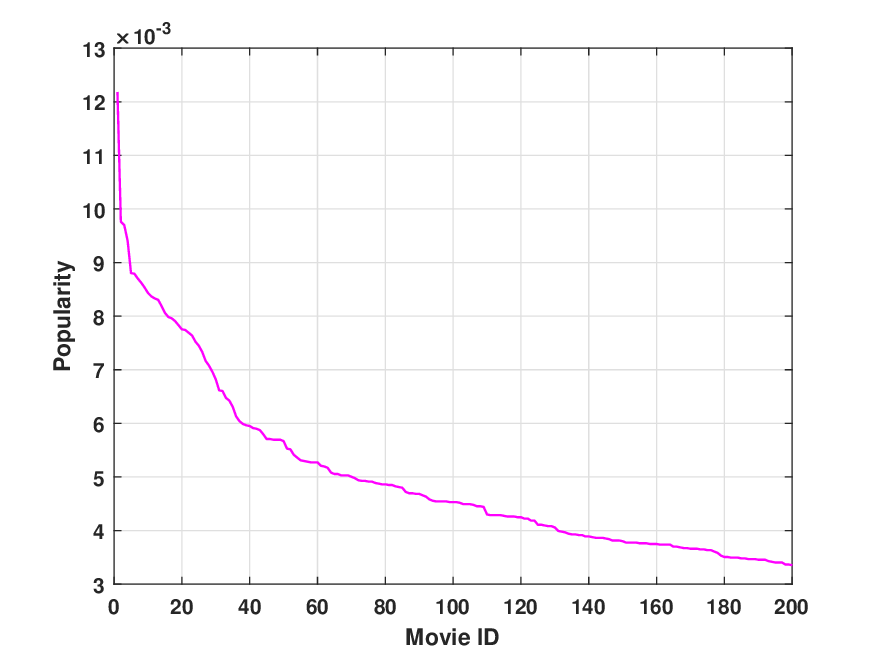} }}%
\caption{LEO satellite beam coverage and location-based UE traffic. (a) Footprint. (b) Traffic. (c) Content demand vector.}
\label{fig4}
\end{figure*}

\begin{figure}[h]%
    \centering
    \subfloat[\centering]{{\includegraphics[width=6 cm, height = 5cm]{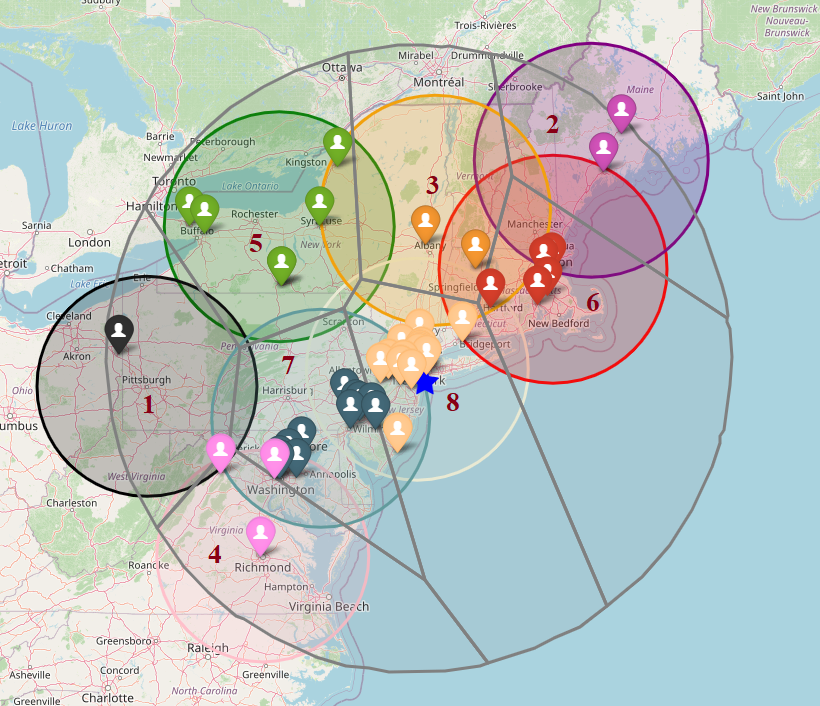} }}%
    \qquad
    \subfloat[\centering]{{\includegraphics[width=6 cm, height = 5cm]{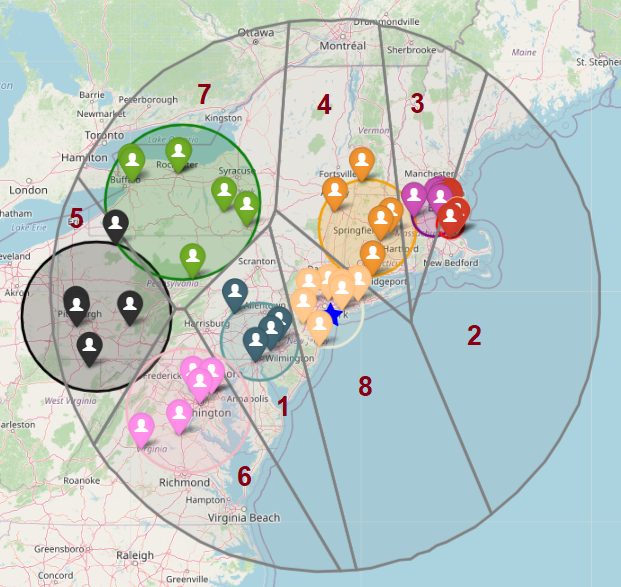} }}%
\caption{ LEO satellite spot beam at time $t$. (a) Fixed beam. (b) Adaptive beam.}
\label{fig5}
\end{figure}

The Starlink LEO satellite 4798 is assumed to be in orbit just above New York (NY) \cite{b45}. The LEO satellite is at an altitude ($H_{s}$) of 550 km just above Earth's surface. The elevation angle ($\epsilon_{\circ}$) of the satellite with the Earth's center is assumed to be $40^\circ$. Based on $H_{s}$ and $\epsilon_{\circ}$, $A_{\sum}$ is about 1.05 million $km^{2}$ with NY as the beam center and a coverage radius ($R_{LEO}$) of $\approx$ 578 km \cite{b46}. The footprint of the LEO satellite is shown in Fig.~\ref{fig4}(a).

\subsection{Content Popularity Based on the Movielens Dataset}
The content popularity is generated from the location-based Movielens dataset, in which 1M movie ratings are provided. The dataset contains UE IDs, UE  locations (ZIP code), movie IDs, movie genres, and rating time, from which we can calculate the distribution of requests in any given time period. We use the ZIP code information to accurately determine the geographic distribution of requests by mapping the ZIP codes with the corresponding latitude and longitude. Since the 1M dataset covers the entire U.S., only UEs falls under the LEO's coverage of the upper part of the U.S. East Coast are considred, shown in Fig.~\ref{fig4}(b). 
After calculating the content popularity $\boldsymbol{L}$ within the covered region, only top 200 movie ID's are taken into account. The most popular movie is indexed as 1, while the least popular is indexed as 200. The popularity of the top 200 movies within LEO satellite beam coverage region is shown in Fig.~\ref{fig4}(c). For each $t$ duration, both the location and the content requested by UEs are randomly changed based on the historical probability distribution.

\subsection{Earth Fixed Beam Duration of LEO Satellite}
In this sub-section, we calculate the elevation angle ($\epsilon_{u}$) of UEs located within the footprint of the Starlink LEO satellite 4798 (see Fig.~\ref{fig5}) for a total connection time of 11 to 12 minutes during one orbital period, taking into account the earth fixed beam scenario. We compute $\epsilon_{u}$ based on the inner product of the LEO satellite's position vector in its orbit and the position vector of the UE, using the earth-centered-earth-fixed coordinate system. Specifically, we use the formula $\epsilon_{u} = \sin^{-1}\left(\frac{\boldsymbol{y}_{u}\cdot(\boldsymbol{y}_{s}-\boldsymbol{y}_{u})}{\lvert\lvert \boldsymbol{y}_{u}\rvert\rvert \lvert\lvert \boldsymbol{y}_{s} - \boldsymbol{y}_{u}\rvert\rvert}\right)$ \cite{b47}, where $\boldsymbol{y}_{s}$ and $\boldsymbol{y}_{u}$ are the position vectors of the satellite and UE, respectively. 
%
% The calculated $\epsilon_{u}$ values for 45 UEs are plotted as a function of time in Fig.~\ref{fig6}.

\begin{figure}[h!]
\centering
% \begin{minipage}{.5\textwidth}
%   \centering
%   \includegraphics[width=1\textwidth]{Result_Fig_5.eps}
%   \caption{Elevation angle with the time variation.}
%   \label{fig6}
% \end{minipage}%
% \qquad
\begin{minipage}{.46\textwidth}
  \centering
  \includegraphics[width=1\textwidth]{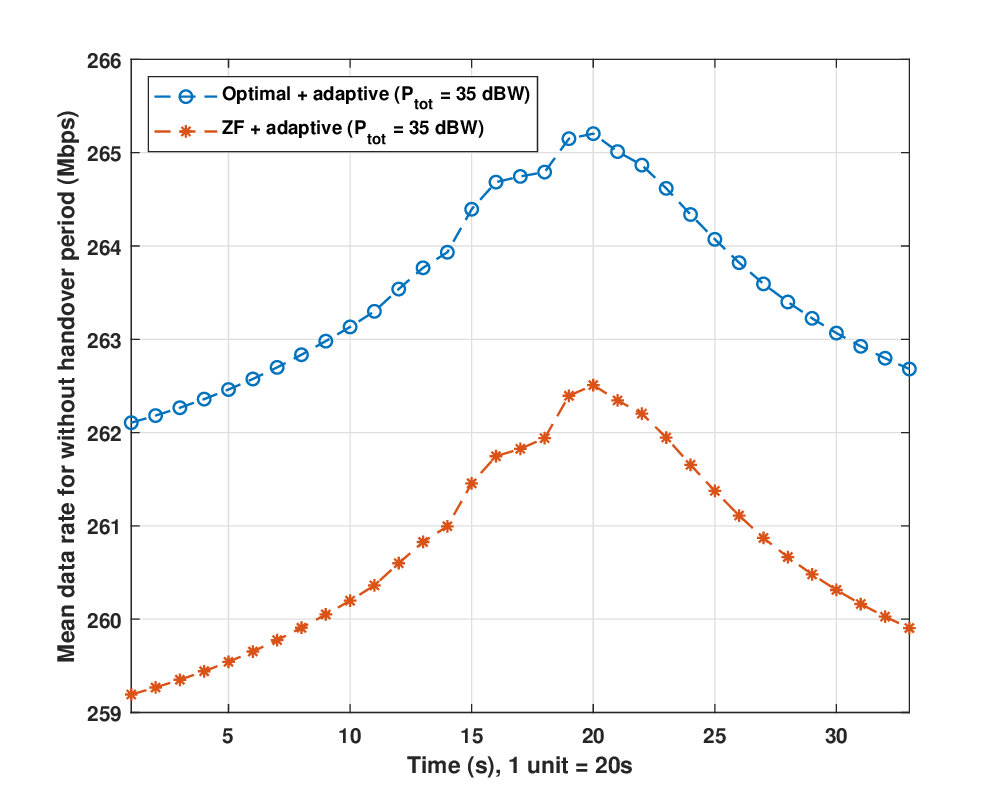}
  \caption{Effective mean data rate of the UEs as a function of time (elevation angle).}
  \label{fig6}
\end{minipage}
\end{figure}

 % Fig.~\ref{fig6} shows that $\epsilon_{u}$ is less than $10^{\circ}$ when the connection duration between LEO satellite and the UE is less than 100 s. As the connection duration increases, $\epsilon_{u}$ also increases, indicating that the LEO satellite is approaching the UEs from the horizon in its orbit. When the connection duration reaches almost half of the total connection time, $\epsilon_{u}$ is about $90^{\circ}$, indicating that the LEO satellite is directly over the beam center. When the connection time increases beyond half the time, $\epsilon_{u}$ indicates that the LEO satellite is moving away from the UEs. In the last minute of the connection time, $\epsilon_{u}$ becomes smaller than $10^{\circ}$, indicating that the LEO satellite is hiding behind the horizon.
 
 Fig.~\ref{fig6} presents the effective mean data rate for optimal and ZF-based precoding designs over time. The figure reveals a noticeable pumping effect in the data rate when the communication time between UEs and LEO satellite reaches its midpoint. This effect occurs because the latched UEs are positioned at an elevation of around $90^{\circ}$ with respect to the LEO satellite during that period.

\subsection{Environment Setup for the CNN-CSI Model}

The training and testing dataset required for the CNN-CSI model is generated by assuming that $U = 45$ UEs are randomly located within the footprint of LEO satellite as shown in  Fig.~\ref{fig5}(b). The channels are time-varying and we apply the Jakes model \cite{b48} to generate the channel matrix. For data generation, we assume that the channel coefficient changes every second. Considering the earth fixed beam scenario, LEO satellite can provide service to all the requesting UEs within its footprint for around 11 minutes as shown in Fig.~\ref{fig6}. Thus, the dataset of size $\mathpzc{D} = 60\times11$ was generated to train, validate, and test the CNN-CSI model. The temporal correlation model used is $\boldsymbol{G}_{t} = \sqrt{\rho} \boldsymbol{G}_{t-1} + \sqrt{1 - \rho} \boldsymbol{E}_{t}$, where $\rho \in [0,1]$ represents the correlation coefficient, and $\boldsymbol{E}_t$ is a time-independent random matrix. A correlation coefficient of 1 indicates complete channel correlation, while a coefficient of 0 indicates channel independence across different time slots.

\begin{figure}
\centering
\begin{minipage}{.46\textwidth}
  \centering
  \includegraphics[width=1\textwidth]{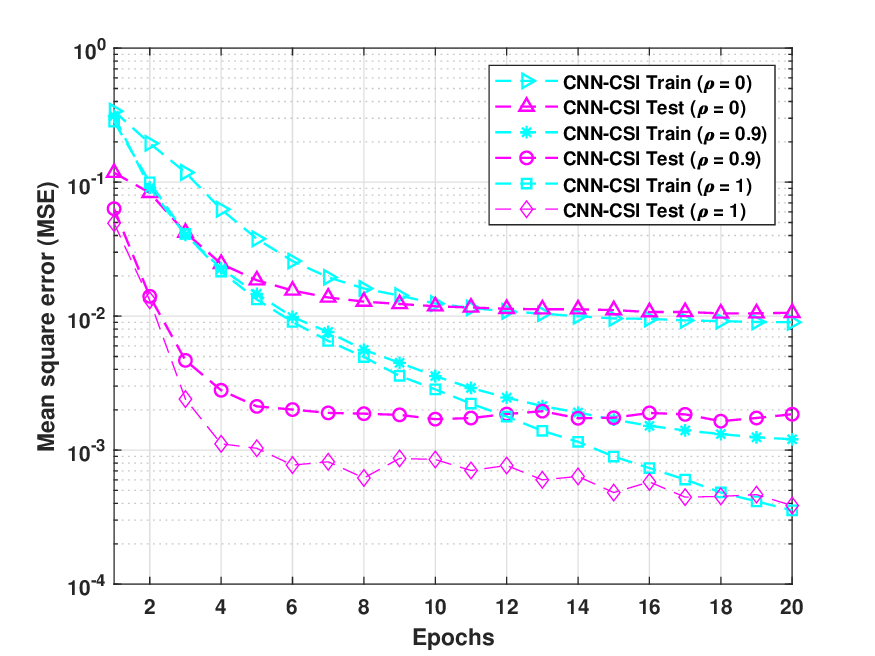}
  \caption{MSE vs. training epochs for CNN-CSI model.}
  \label{fig7}
\end{minipage}%
\qquad
\begin{minipage}{.46\textwidth}
  \centering
  \includegraphics[width=1\textwidth]{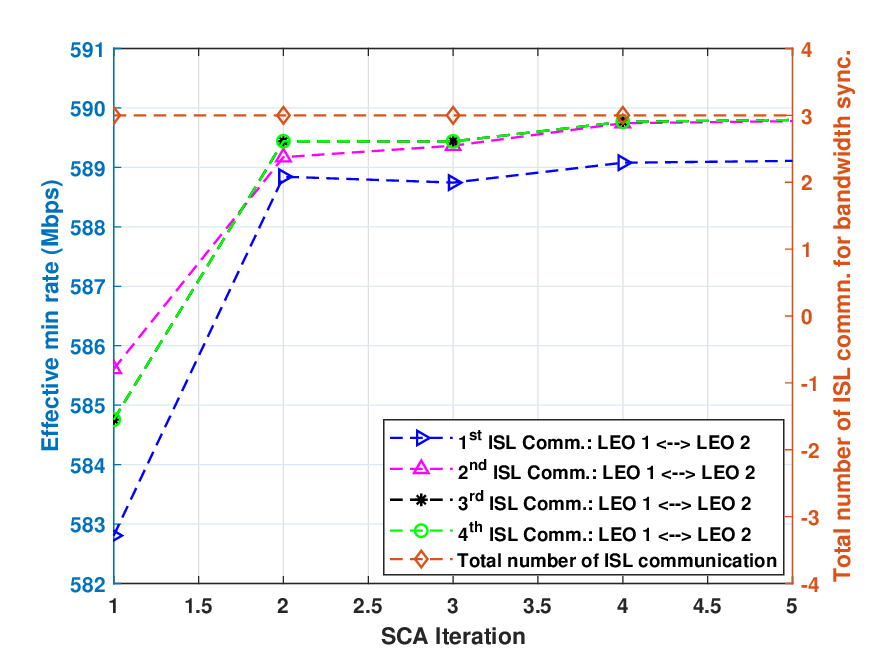}
  \caption{Convergence of the DC-HO algorithm.}
  \label{fig8}
\end{minipage}
\end{figure}

As to the CNN-CSI model, we employ $l_c = 3$ layers where the number of filters and the filter size of corresponding layers are $\{16, 8, 4\}$ and $\{3\times3, 3\times3, 3\times3\}$, respectively. Other parameters are summarized in Table II.
Fig.~\ref{fig7} illustrates the accuracy of the CNN-CSI model in terms of mean square error (MSE) for different training epochs. The training and test datasets were generated with different correlation coefficient values, denoted $\rho$. As shown in Table II, the total number of training, validation, and test dataset are 528, 130, and 1, respectively. 

% This totals 659 datasets, aligning with the communication window observed in Figure~\ref{fig6}.}

Since we are interested in predicting the channel coefficients of LEO satellite 2 during the handover time slot, only one test dataset is required. This means that, after the CNN-CSI model is well-trained, the data from the $659^{th}$ time slot is used as input to predict the next time slot's channel coefficient for LEO satellite 2.  Specifically, we considered $\rho$ values of 1, 0.9, and 0, for which the corresponding MSE values for training the model were 0.000385, 0.00185, and 0.0105, respectively. As from these results, it can be concluded that the prediction accuracy of the model CNN-CSI is higher when the temporal correlation of the channel coefficient is high and vice-versa. Additionally, the test error is slightly lower than the training error for a smaller number of training epochs, primarily due to the significantly smaller size of the test dataset and the model being inadequately trained during these early epochs. However, as the number of training epochs increases, both the training and test errors converge, and the MSE decreases, demonstrating that our CNN-CSI model is well-trained for datasets with different $\rho$ values.

\subsection{Performance Evaluations }

\begin{table}[h!]
\centering
% \begin{minipage}{.45\linewidth}
\centering
\caption{CNN-CSI Training Parameters}
\begin{tabular}{|p{3.5cm}||p{2cm}|}
\hline
\textbf{Parameters} & \textbf{Value} \\
\hline
2D data size & 45 $\times$ 2 \\
 Total training dataset & 528 \\
 Total validation dataset & 130 \\
 Test dataset & 1 \\
 Maximum training epochs, $E_{max}$ & 20 \\
 Dropout &   0.4  \\
 Optimizer & Adam \\
Activation functions & ReLU \& Tanh \\
 Loss function & MSE \\
 Learning rate, $\alpha$ & 0.0001\\ 
\hline
\end{tabular}
% \end{minipage}%
\end{table}

\begin{table}[h!]
\centering
% \begin{minipage}{.45\linewidth}
\centering
\caption{System and Channel Parameters}
\begin{tabular}{|p{4cm}||p{2.75cm}|}
\hline
\textbf{Parameters} & \textbf{Value} \\
\hline
Number of UEs, $U$ & 45  \\
Parallel data streams, $N$ &   4  \\
Number of spot beams, $M$ &   8  \\
Carrier frequency, $f_c$ & 2 GHz \\
Bandwidth, $B$ & 100 MHz \\
Rician factor, $\kappa_u$ & 10 dB \\
Spot beam gain, $G_m$ & U [18 dBi, 36 dBi]  \\
Spot beam radius, $r_m$ & U [25 km, 200 km]  \\
AoD, $\varphi_u$ & U [-0.5, 0.5] \\
UE's elevation angles, $\epsilon_{u}$ & U [$40^{\circ}$, $90^{\circ}$]\\ 
Transmit power, $P_{\sum}$ & 30 dBW -- 40 dBW \\
QoS, $R_{req}$ & $0.25\times b_{a,m}$ \\
Backhaul rate, $R_{BH}$ & 0.25 Gbps -- 1 Gbps\\ 
Size of movies files, $q$ & U [0.3 Gb, 1.8 Gb]\\
Cache storage capacity, $C$  & 215 Gb \\
Noise spectral density, $N_{0}$ & -203.9772 dBW/Hz \\
Adaptive spot beam radius, $r_{m}$ & U [25, 200] km \\
Fixed spot beam radius &  200 km \\
Earth's radius &  6371 km \\
\hline
\end{tabular}
% \end{minipage}
\end{table}

In this part, we conduct the numerical results considering a scenario where LEO satellite has a total of 8 spot beams serving 45 UEs. The UEs are randomly distributed in the coverage area as shown in Fig.~\ref{fig5}. It is assumed that each spot beam is capable of transmitting $N = 4$ parallel data streams. 
% It is assumed that LEO satellite is able to generate 8 spot beams using the beamforming technology. 
The coverage area and the number of UEs within each spot beam is calculated using Algorithm~\ref{algorithm:1}. The spot beam with the lowest number of UEs is marked as 1, while the spot beam with the highest number of UEs is marked as 8. 
% The number 8 is shown in the spot beam where the beam center is located. 
Fig.~\ref{fig5}(a) displays a fixed-spot beam, while Fig.~\ref{fig5}(b) demonstrates a steerable adaptive spot beam. We adopt the LTE specifications \cite{b49}, where one c.u. lasts one symbol duration, which is equal
to 66.7 $\mu s$, and one block duration comprises 300 c.u. The LEO satellite is assumed to spend 1 c.u. to solve one convex optimization problem, resulting in $M$ c.u. for solving each proposed algorithm \cite{b50}. The system and channel parameters used in the simulations are summarized in Table III. The simulation results are averaged over the 100 random channel realizations. 
% \textcolor{blue}{In these simulations, we haven't included a comparison with the terrestrial multicasting solution, as satellite systems are designed to complement terrestrial networks.} 
We compare the proposed framework with the following references:

\begin{itemize}
\item \emph{Baseline 1:} The precoding vectors are designed based on the ZF-based approach.
\item \emph{Baseline 2:} The optimal approach is used to calculate the precoding vectors for the combined channel coefficients of the participating LEO satellites in the CC-HO method. In this approach, both satellites estimate the CSI via pilot transmission. As a result, the portion of time responsible for data transmission changes to $(1 - \frac{{2\tau_{csi}+\tau_{pro}}}{{\tau_{slot}}})$.
\item \emph{Baseline 3:} The channel estimation is similar in \emph{Baseline 2}, except that the precoding vectors are design based on the ZF method.
% based approach is used to calculate the precoding vectors for the combined channel coefficients of the participating LEO satellites in the CC-HO method. In this approach, the downlink channel coefficient for LEO satellite 2 is determined directly by the LEO satellite 1 without relying on the prediction model. As a result, the portion of time responsible for data transmission changes to $(1 - \frac{{2\tau_{csi}+\tau_{pro}}}{{\tau_{block}}})$.
\end{itemize}
The comparison with the terrestrial multicasting solution is not considered as satellite systems is design to complement the terrestrial networks.

\subsubsection{Convergence of the Proposed Iterative Algorithm}

To demonstrate the convergence of our proposed algorithms, Fig.~\ref{fig8} presents the objective function of the iterative algorithms for baseline 2 with a focus on min rate maximization (maximum content delivery delay minimization). 
% The x-axis represents the SCA iterations, while the y-axis represents the objective function values. 
The total transmit power of LEO satellite 1 and LEO satellite 2 is considered 25 dBW each. It is evident from the plot that the iterative algorithms exhibit rapid convergence to the sub-optimal values, requiring fewer than 5 iterations and 3 ISL communications for the DC-HO scenario.

\subsubsection{Without Handover Scenario}

\begin{figure}
\centering
\begin{minipage}{.46\textwidth}
  \centering
  \includegraphics[width=1\textwidth]{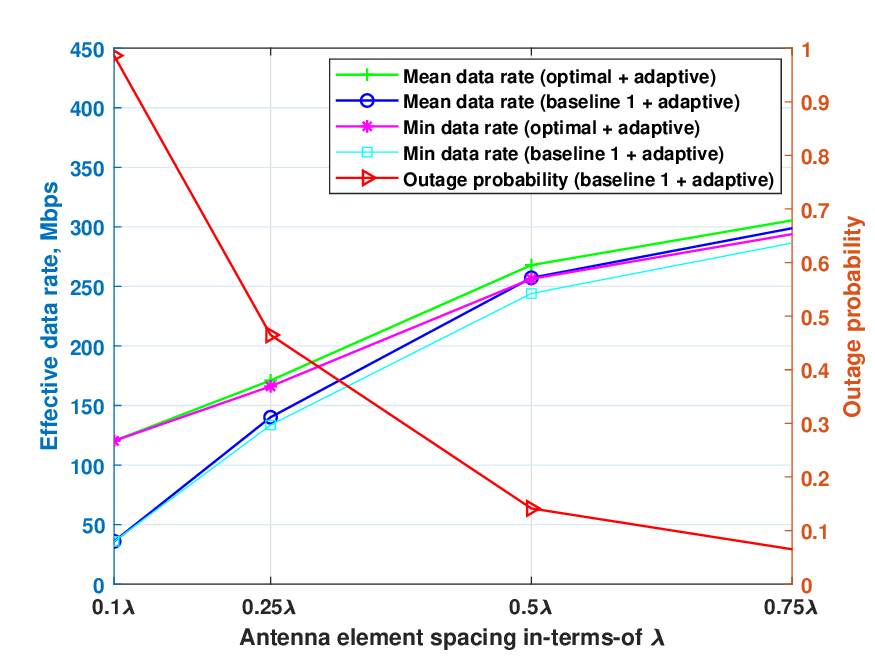}
  \caption{Effective mean data rate and the outage probability versus the LEO satellite array antenna element spacing ($d_{ant}$).}
  \label{fig9}
\end{minipage}%
\qquad
\begin{minipage}{.46\textwidth}
  \centering
  \includegraphics[width=1\textwidth]{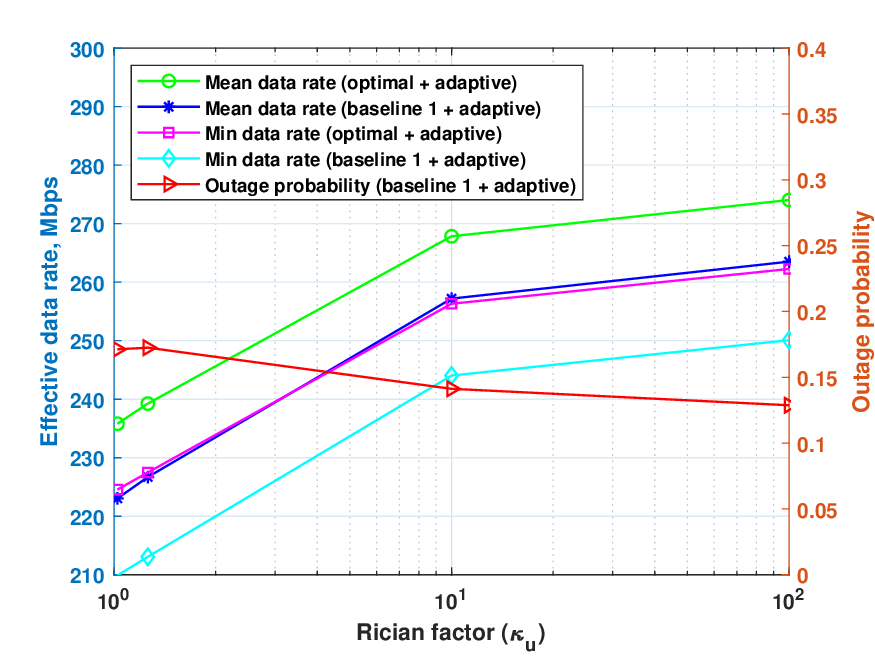}
  \caption{Effective mean data rate and the outage probability versus the Rician factor ($\kappa_u$).}
  \label{fig10}
\end{minipage}
\end{figure}

\begin{figure}
\centering
\begin{minipage}{.46\textwidth}
  \centering
  \includegraphics[width=1\textwidth]{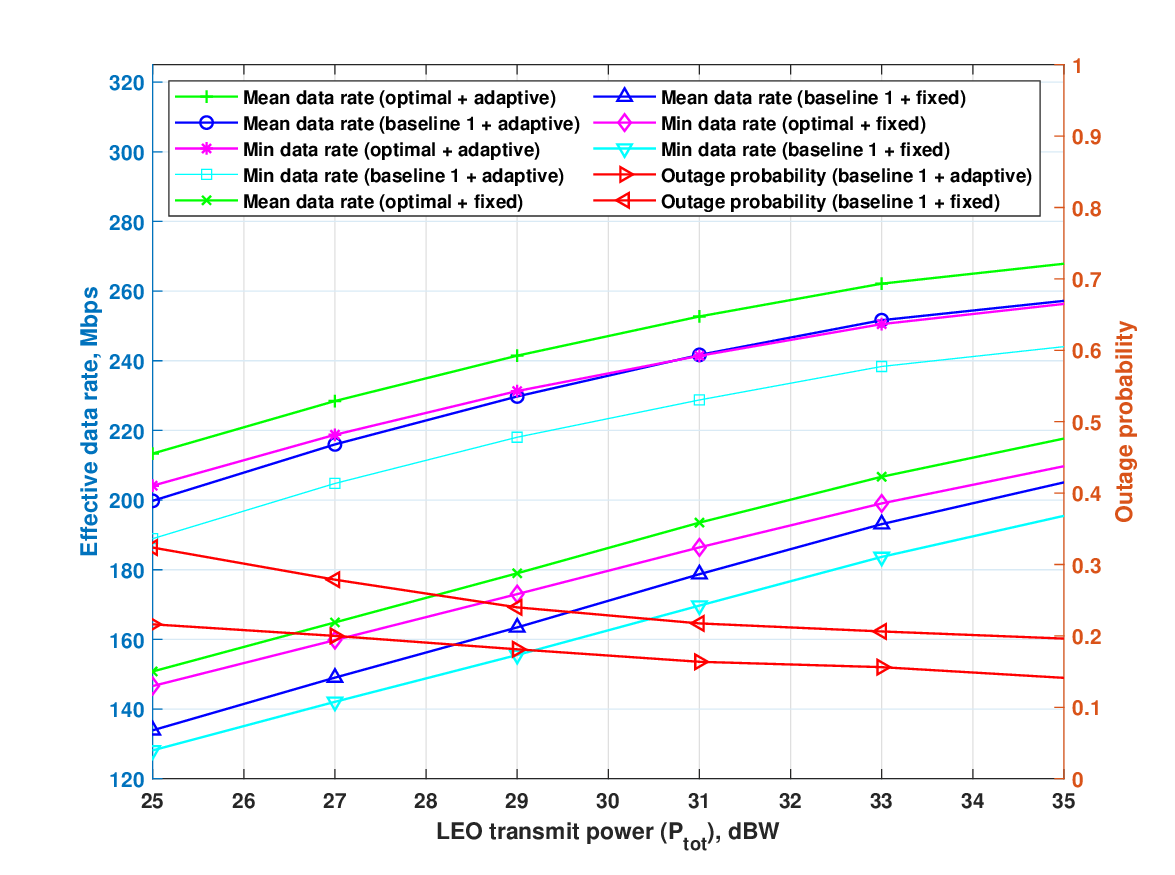}
  \caption{Effective data rate and the outage probability versus the total transmit power of the LEO satellite.}
  \label{fig11}
\end{minipage}%
\qquad
\begin{minipage}{.46\textwidth}
  \centering
  \includegraphics[width=1\textwidth]{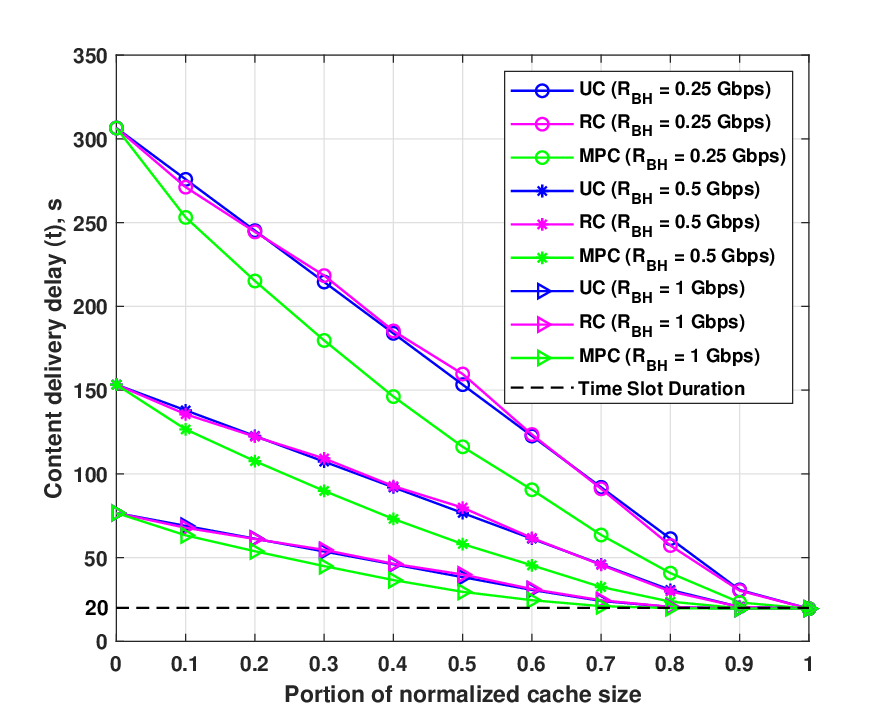}
  \caption{Content delivery delay versus the portion of normalized cache size for different generic caching scenarios.}
  \label{fig12}
\end{minipage}
\end{figure}

Fig.~\ref{fig9} portrays the mean/min data rate and outage probability between the proposed optimal design and baseline 1 for various values of LEO satellite antenna element spacing, i.e., $d_{ant}$. The optimal design consistently achieves higher mean/min data rates than baseline 1. At a lower $d_{ant}$ of $0.1\lambda$, the optimal design achieves about 3.35 times the mean/min data rate of baseline 1. In contrast, at an $d_{ant}$ of $0.75\lambda$, the optimal design achieves a mean/min data rate about 1.02 times that of baseline 1. Baseline 1 performs poorly at lower $d_{ant}$ due to more sidelobes and lower spot beam directivity, leading to significant outage scenarios caused by channel correlation. In contrast, the optimal design experiences no outages. The figure reveals that increasing the $d_{ant}$ results in higher data rates for both designs. However, increasing $d_{ant}$ beyond $0.5\lambda$ raises the risk of grating lobes, making $d_{ant}$ of $0.5\lambda$ desirable. At this spacing, the mean data rate for the optimal design reaches around 267 Mbps, while baseline 1 achieves approximately 257 Mbps. 

Fig.~\ref{fig10} depicts the mean/min data rate and outage probability as a function of $\kappa_u$ for both the optimal and baseline 1 approaches considering the adaptive beam scenario. 
% The y-axis of Fig.~\ref{fig10} is the same as in Fig.~\ref{fig9}. 
For this result, $d_{ant}$ is considered to be $0.5\lambda$ and $P_{\sum}$ to be 35 dBW. The $\kappa_u$ is varied between 1.0233 (0.1 dB) and 100 (20 dB). From the figure, it can be seen that the data rate (mean/min) increases as the value of $\kappa_u$ increases and that the data rate obtained with the optimal precoding design is higher than that of baseline 1 regardless of the $\kappa_u$ value. When $\kappa_u$ is 0.1 dB, the outage probability due to baseline 1 is about 17\% and when $\kappa_u$ is 100, the outage probability due to the baseline 1 design is 12\%, while there is no outage scenario due to the optimal precoding design for the given $R_{req}$. Since the rates and outage probability are not significantly different between $\kappa_u$ of 10 (10 dB) and 100 (20 dB), $\kappa_u$ of 10 dB is considered, which is a realistic assumption.

Fig.~\ref{fig11} demonstrates the mean/min data rate and outage probability as a function of $P_{\sum}$, with $d_{ant} = 0.5\lambda$, $\kappa_u  = 10$dB. The figure demonstrates that the mean/min data rate achieved by the adaptive beam scenario is at least 1.22 times higher than that of the fixed beam scenario, regardless of the precoding approaches.  In the fixed beam scenario, all the spot beams have equal radius of $200$ km, whereas the spot beam's radius in the adaptive beam scenario varies between 25 km and 200 km. The adaptive beam scenario, benefiting from improved beam directivity, outperforms the fixed-beam scenarios in terms of both data rate and outage probability. Moreover, optimal precoding achieves a significantly higher mean/min data rate than baseline 1 because it effectively eliminates intra-spot beam interference. There is no outage scenario in the optimal precoding-based design, while there are high outage probability in the baseline approach. 
When $P_{\sum}$ is increased, the outage probability decreases, and therefore the data rate improves comparatively more in the baseline approach than in the optimal approach. However, the increase in data rate tends to saturate when the power increase exceeds a certain limit because the inter-spot beam interference increases due to the power leakage from the neighboring beams.

Fig.~\ref{fig12} illustrates the delay as a function of normalized cache capacity for the caching models such as MPC ($\mu_{k} \in \{0,1\}$), UC ($\mu_{k} \in [0,1]$), and RC ($\mu_{k} \in \{0,1\}$). From the figure, it can be seen that the delivery time is lower in the MPC approach compared to the UC and RC methods for small normalized cache sizes, and RC almost approaches the UC method due to the averaging of a large number of channel realizations. When the normalized cache size increases, both UC and RC approach the performance of the MPC. It can also be seen from the figure that $R_{BH}$ also significantly affects the delivery latency for the different cache size values. 

\subsubsection{Handover Scenario}

\begin{figure}[ht]
\centerline{\includegraphics[width=0.5\textwidth]{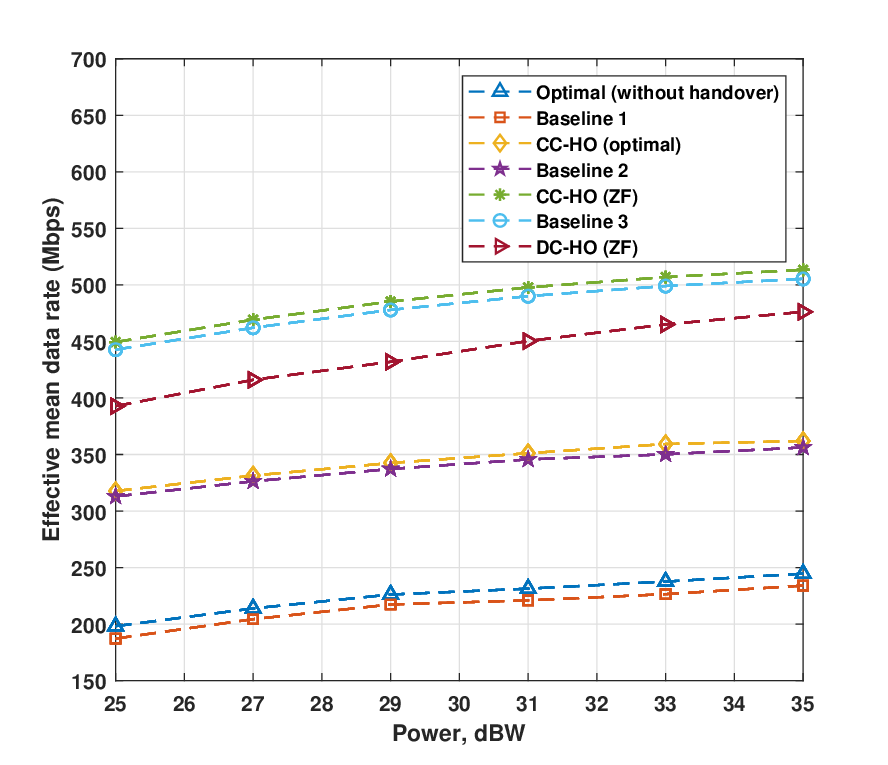}}
\caption{Effective mean data rate versus
the total transmit power of the LEO satellites for the HO scenario.}
\label{fig13}
\end{figure}

Fig.~\ref{fig13} shows the relationship between the transmit power and the effective mean data rate during the HO process. The figure reveals that regardless of the precoding scheme used, the effective mean data rate during HO is consistently 1.5 times higher than the without HO period. Comparing the proposed approaches with the baselines, the ZF precoding design-based CC-HO approach outperforms baseline 3, and the optimal precoding design-based CC-HO approach outperforms baseline 2. This is because the proposed CC-HO approach estimates the CSI using a prediction model, while baseline 2 and baseline 3 estimate the CSI via pilot transmission. Despite utilizing ZF-based precoding design in the CC-HO scenario, it outperforms other approaches due to the combined consideration of downlink channel coefficients and evaluation of the precoding vectors based on the combined channel coefficients, which enhances spatial diversity. 
% Consequently, the correlation among UEs' channel coefficients is reduced, resulting in improved performance for the ZF-based approach. It effectively mitigates intra-spot beam interference, making it superior to the proposed optimal approach for the CC-HO. 
The ZF-based CC-HO outperforms ZF-based DC-HO, even though both employ ZF-based precoding. This is because the former allows for full control in the design of the precoding vector and bandwidth allocation. In contrast, the latter requires synchronization of resources between the LEO satellites involved in the HO, which might not always be guaranteed.

\subsection{Impact of Imperfect CSI}
\begin{figure}[ht!]
\centerline{\includegraphics[width=0.5\textwidth]{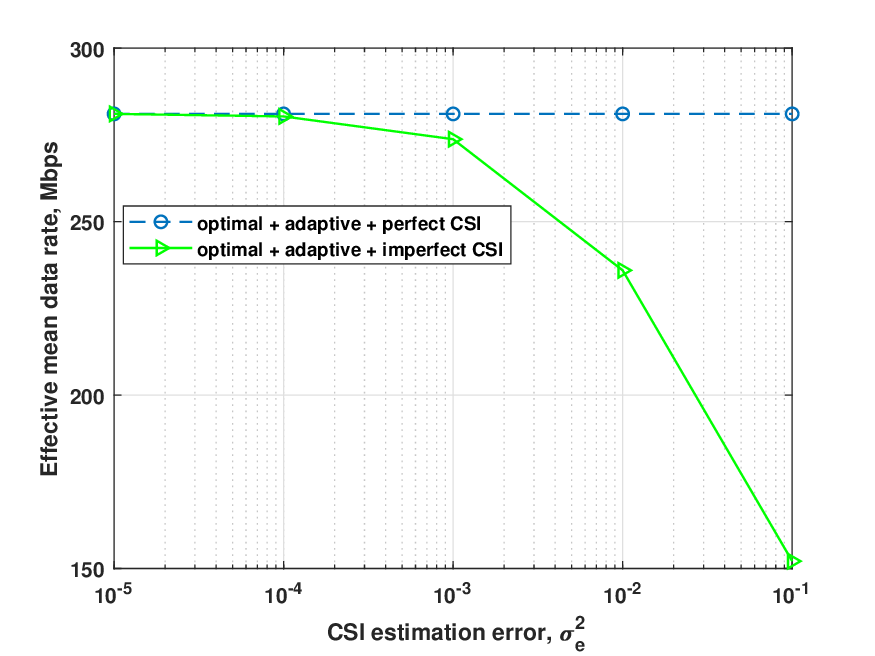}}
\caption{Effective mean data rate versus
error variance ($\sigma_{e}^{2}$) on the imperfect CSI for non-HO scenario for a multi-spot beam multicasting LEO satellite system ($P_{\sum} = 35$ dBW).}
\label{fig14}
\end{figure}
In previous sections, the proposed framework assumes perfect CSI at the satellite. In realistic conditions, the satellite operates based on the imperfect channel estimation $\hat{\boldsymbol{h}}_{u,k,a,m} = \boldsymbol{h}_{u,k,a,m} + \boldsymbol{e}$, where $\boldsymbol{h}_{u,k,a,m}$ is the true channel and $\boldsymbol{e}$ is the estimation error that is independent from the true channel and follows $\mathcal{CN}(0,\boldsymbol{I}\sigma^2_e)$. Since the precoding vectors are designed based on the estimated channels, the SINR in this case equals to:
 % \begin{equation}
 % \small
$\frac{|\boldsymbol{h}_{u,k,a,m}^H \boldsymbol{w}_{k,a,m}|^2}{\underset{k' \neq k}{\sum}|{\boldsymbol{h}_{u,k,a,m}^{H}\boldsymbol{w}_{k',a,m}}|^2+ \underset{ \forall k'}{\sum} \sigma^2_e \|\boldsymbol{w}_{k',a,m}\|^2 + \hat{I}_{agg}b_{a,m}}, \notag
$
% \end{equation}
%
where the summation of $k'$ in the denominator is over $\mathcal{K}_{a,m}$.
Optimal bandwidth allocation and precoding vectors design under imperfect CSI can be obtained similarly in Section~\ref{sec:jointframework}, with one modification of adding $\sum_{ \forall k'} \sigma^2_e \|\boldsymbol{w}_{k',a,m}\|^2$ to the RHS of \eqref{eq18}. 
%
% Therefore, to illustrate the impact of imperfect CSI on the data rate, the effective mean data rate for varying values of the estimation error, i.e., $\epsilon_{u} \sim$ $\mathcal{CN}(0,\sigma_{\epsilon_{u}}^2)$,  associated with the imperfect CSI of the user is shown in Fig. 14. For ease of discussion, we consider a scenario of a constant error variance, i.e., $\sigma_{\epsilon_{u}}^2 = \sigma_{\epsilon}^2$, for all users.}

Fig.~\ref{fig14} presents the impact of imperfect CSI on the proposed optimal precoding design. The robustness of our design is demonstrated via a close performance to the perfect CSI case for estimation error up to $10^{-3}$. When the estimation error is large, the achievable rate dramatically degrades as expected.

\section{Conclusion}
In this paper, we have proposed a FLARE-LEO framework that effectively exploits the fully flexible regenerative payload capability of LEO satellites via joint design of spot beam coverage, adaptive beamforming, caching, multiuser precoding and dynamic bandwidth allocation using realistic system parameters.
In addition, we proposed innovative handover architectures that consider computational capability and overhead requirements. Using numerical results, we demonstrated that our adaptive beamforming design outperforms the fixed beam design in terms of both data rate and outage probability. We also showed that collaboration between two LEO satellites during the HO period significantly boosts the system performance. In general, the optimal precoding design outperforms the ZF-based precoding design, resulting in both an improvement in data rate and a reduction in the content delivery latency. Furthermore, we have shown that the MPC-based caching method performs better than caching strategies based on RC and UC and significantly improves the average content delivery latency for content delivery compared to a scenario without caching.

From the outcomes of this work, a promising topic is to consider a network of LEO satellites in which the beams' coverage management should be jointly designed between multiple LEO satellites, taking into account imperfect CSI conditions. Another interesting topic is to study the handover in LEO satellties when the UEs are equipped with multiple antennas. In this case, the UEs can establish connection with multiple LEO satellites to improve the effective data rate.

\begin{appendices}
\section{Proof of Proposition 1}

Consider a function $\mathrm{g}(u,v)$ in $\mathbb{R}^{2}_{+}$. The Hessian of $\mathrm{g}(u,v)= u\log(1+v/u)$ is given as follow:

 \begin{equation}
  \mathcal{H}_\mathrm{g} =
  \begin{bmatrix}
    \dfrac{-v^2}{u(u+v)^2} & \dfrac{v}{(u+v)^2}  \\
    \dfrac{v}{(u+v)^2} & -\dfrac{u}{(u+v)^2} 
  \end{bmatrix}
  \label{eq:myeqn}
\end{equation} 

For arbitrary vector $\mathbf{x}=[p \,\, q ]^{T}$, we can calculate $\mathbf{x}^T \mathcal{H}_\mathrm{g} \mathbf{x}
= -\frac{(pv-qu)^2}{u^2(u+v)^2}$, which is always non-positive. 
This implies that the function $u\log(1+v/u)$ is concave in its supports. From $\eqref{eq31}$, we can write $R_{1,k,a,m}^{DC,ZF}= \Phi\mathrm{g}(\hat{I}_{agg}^{HO}{b}_{1,a,m},{x}_{1,k,a,m})/\hat{I}_{agg}^{HO}$, which completes the proof of Proposition 1.
\end{appendices}

\balance

\end{document}